\documentclass[letterpaper,english,twocolumn, aps]{revtex4-1}
\usepackage{color}
\usepackage{amsmath}
\usepackage{amssymb}
\usepackage{graphicx}

\makeatletter

\pdfpageheight\paperheight
\pdfpagewidth\paperwidth

\newcommand{\rev}[1]{\textcolor{black}{#1}}

\newcommand{\sigmaeff}{\sigma_{\mbox{\tiny eff}}}
\newcommand{\thetaads}{\theta_{\mbox{\tiny ads}}}

\makeatother

\usepackage{babel}
\begin{document}
\title{Polymer brushes with reversibly tunable grafting density}
\author{Leonid I. Klushin}
\affiliation{Department of Physics, American University of Beirut. P. O. Box 11-0236,
Beirut 1107 2020, Lebanon}
\affiliation{Institute of Macromolecular Compounds, Russian Academy of Sciences. 31
Bolshoy pr, 199004 Saint Petersburg, Russia}
\author{Alexander M. Skvortsov}

\affiliation{Chemical-Pharmaceutical University. Professora Popova 14, 197022 St. Petersburg,
Russia}
\author{Alexey A. Polotsky}
\affiliation{Institute of Macromolecular Compounds, Russian Academy of Sciences. 31
Bolshoy pr, 199004 Saint Petersburg, Russia}
\author{Anna S. Ivanova}
\affiliation{Institute of Macromolecular Compounds, Russian Academy of Sciences. 31
Bolshoy pr, 199004 Saint Petersburg, Russia}
\author{Friederike Schmid}
\affiliation{Institut f\"ur Physik, Johannes Gutenberg-Universit\"at Mainz, Staudingerweg
7, D-55099 Mainz, Germany}
\begin{abstract}

We propose a novel class of responsive polymer brushes, where
the effective grafting density can be controlled by external stimuli.
This is achieved by using \rev{end-grafted polymer chains}
that have an affinity to the substrate. For sufficiently strong surface
interactions, a fraction of chains condenses into a near-surface layer,
while the remaining ones form the outer brush. The dense layer and
the more tenuous outer brush can be seen as coexisting microphases.
The effective grafting density of the outer brush is controlled by
the adsorption strength and can be changed reversibly and in a controlled
way as a response to changes in environmental parameters. The effect
is demonstrated by numerical SCF calculations and analyzed by scaling
arguments. Since the thickness of the denser layer is about a few
monomer sizes, its capacity to form a microphase is limited by the
product of the brush chain length and the grafting density. We explore
the range of chain lengths and grafting densities where the effect
is most pronounced. In this range, the SCF studies suggest that
individual chains inside the brush show large rapid fluctuations between
two states that are separated by only a small free energy barrier.
The behavior of the brush as a whole, however, does not reflect these
large fluctuations, and the effective grafting density varies smoothly
as a function of the control parameters.

\end{abstract}
\maketitle

\section{Introduction}

Polymer brushes are commonly used for permanent surface modification
to mediate the stability of colloidal dispersions, provide anti-fouling
properties, and protect the system from degradation \citep{Currie:2003,Ayres:2010,Urban:2011,Jaquet:2013,Motornov:2003}.
Brushes can also act as smart stimuli-responsive materials that change
surface wetting properties reversibly or act as sensors \citep{Cohen-Stuart:2010,Qi:2015,Chen:2010,Gupta:2008,Merlitz:2009}.

Under good solvent conditions, the physical properties of brushes
are determined by two key \rev{parameters}: the chain length and the grafting
density. Together, they determine the thickness of the brush layer
and the strength of the repulsive forces that the brush exerts on
objects approaching the surface. In general, both the chain length
and the grafting density are set at the stage of the brush synthesis
and cannot be changed thereafter. This is also the situation which
is considered in most theoretical studies.


The aim of the present paper is to extend this concept. We propose
to use brushes formed by \rev{end-grafted} adsorption-active
chains, in which case properties are affected by the short-range adsorption
interactions between the substrate and the chain units. The properties
of such brushes are studied by self-sonsistent field (SCF) calculations
and a scaling analysis. \rev{We predict that over a wide
range of adsorption strengths, part of the brush chains are almost
completely laid out on the surface while the remaining chains form
a brush with a reduced grafting density. Hence it is possible to reversibly
control the effective brush grafting density and, therefore, its properties,
by changing the strength of the adsorption potential. Experimentally,
the adsorption strength can be changed in two ways: 1) by changing
the composition of a mixed solvent (this can result in a very strong
variation of the adsorption parameter), and 2) by changing the temperature
of a mixed solvent which would result in a finer tuning). A large
array of mixed solvents adjusted for specific polymer-substrate pairs
was developed in the context of liquid chromatography
studies~\cite{Pasch:2014} }

\rev{ A brush is often considered as a semi-dilute polymer system; it
is natural to expect some similarities between the behavior of the
brush composed by adsorption-active chains and the adsorption from a
semi-dilute solution. In both cases, the attraction to the substrate
competes with a local accumulation of monomers leading to an increase
in steric repulsion and eventually to saturation. In the case of
adsorption from solution, bound chains form a dense proximal layer
whereby they are in contact with the substrate, but also a tenuous
more distant layer composed by tails~\cite{Fleer:1993}.  In the case
of adsorption-active brushes a new factor comes into play: All chains
are permanently attached to the surface even when they are not bound
by adsorption.  Hence, two scenarios are conceivable: 1) a part of
each chain starting from the grafted end is adsorbed while the rest of
the chain forms a tail, and 2) a certain fraction of chains is fully
adsorbed, and the remaining ones are desorbed and form the outer
brush. We demonstrate below that the second scenario applies.}

\section{Model and method}

We consider a polymer monodisperse brush made of linear flexible macromolecules
grafted at one end onto a solid planar substrate. A polymer chain
is composed of $N$ identical monomer units, the chains are grafted
onto the surface at the grafting density $\sigma$, defined as the
number of grafted polymer chains per unit surface area. The surface
is assumed to be attractive to all polymer chains, the monomer-surface
attraction is characterized by the adsorption energy $-\varepsilon$,
$\varepsilon>0$. The brush is immersed into an athermal solvent;
in terms of Flory-Huggins interaction parameter $\chi$ this corresponds
to $\chi=0$. To calculate the system's partition function and its
various properties, we use the the Scheutjens--Fleer self-consistent
field (SF-SCF) method. The SF-SCF method and its modifications for
the study of polymer brushes of various types have been repeatedly
described in the literature and can be found, for example, in \citep{Fleer:1993}.
The SF-SCF approach uses a lattice, which facilitates to account for
the volume of all molecular components, and also takes into account
the symmetry of the problem under consideration. Polymer chains are,
therefore modeled as walks on the simple cubic lattice. The lattice
cell size is equal to the size of a monomer unit, each lattice site
can be occupied either by a monomer unit by or a solvent molecule. The lattice
sites are organized in a planar layers, each layer is referred to
with a coordinate $z$ normal to the grafting plane. Within a layer
with fixed $z$, i.e. along $x$ and $y$ axes, the volume fractions
of the monomeric components and the self-consistent potential are
taken as uniform; hence, we use a one-gradient version of the SF-SCF
method for planar geometry. A monomer unit in the first lattice layer
adjacent to the surface has a contact with the surface and acquires
an additional energy gain $-\varepsilon$\@. More details about the
implementation of the SF-SCF method is given in Appendix A. In
order to obtain a physical understanding of the effects observed in
the SCF calculations and characterize the crossover between different
regimes, we complement them with a scaling analysis using ''blob''
concepts as outlined in Refs.~\cite{Halperin:1994,Rubinstein_book}.

Throughout this paper, energies are given in units of $k_{B}T$
and lengths in units of the statistical segment length or monomer
size $a$, corresponding to the lattice cell size in the Scheutjens-Fleer
method.

\section{Results of the SCF Calculations}

\subsection{Adsorption regimes in the brush}

\label{sec:regimes}

\begin{figure}
\begin{centering}
\includegraphics[width=8cm]{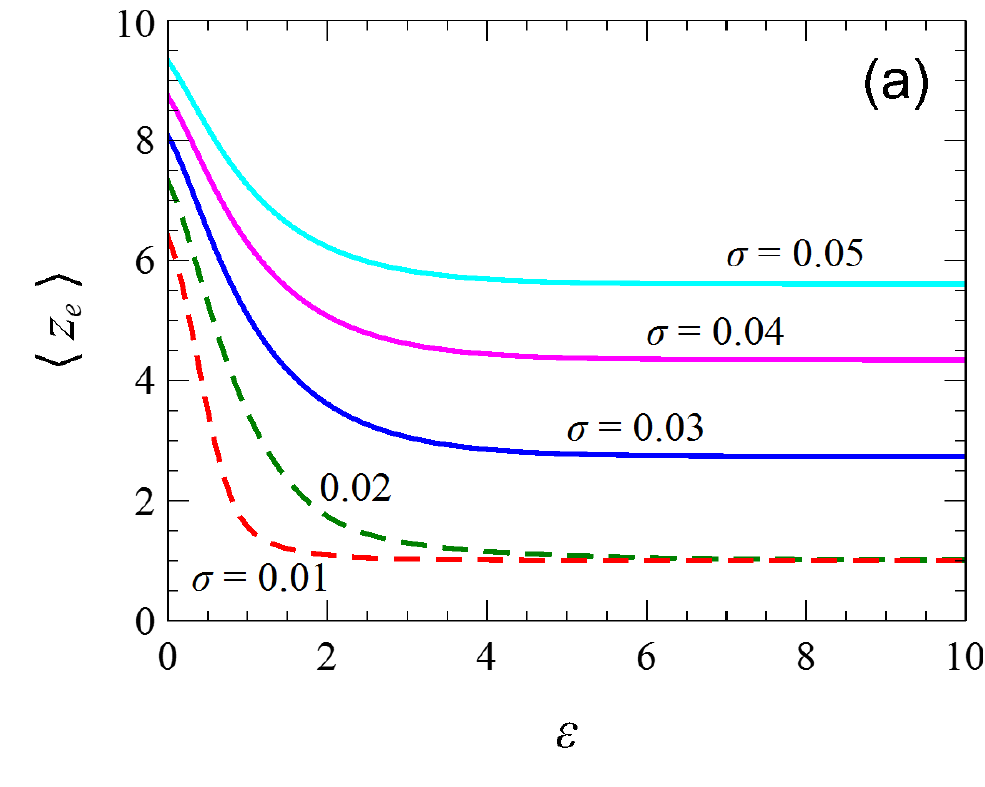} \includegraphics[width=8cm]{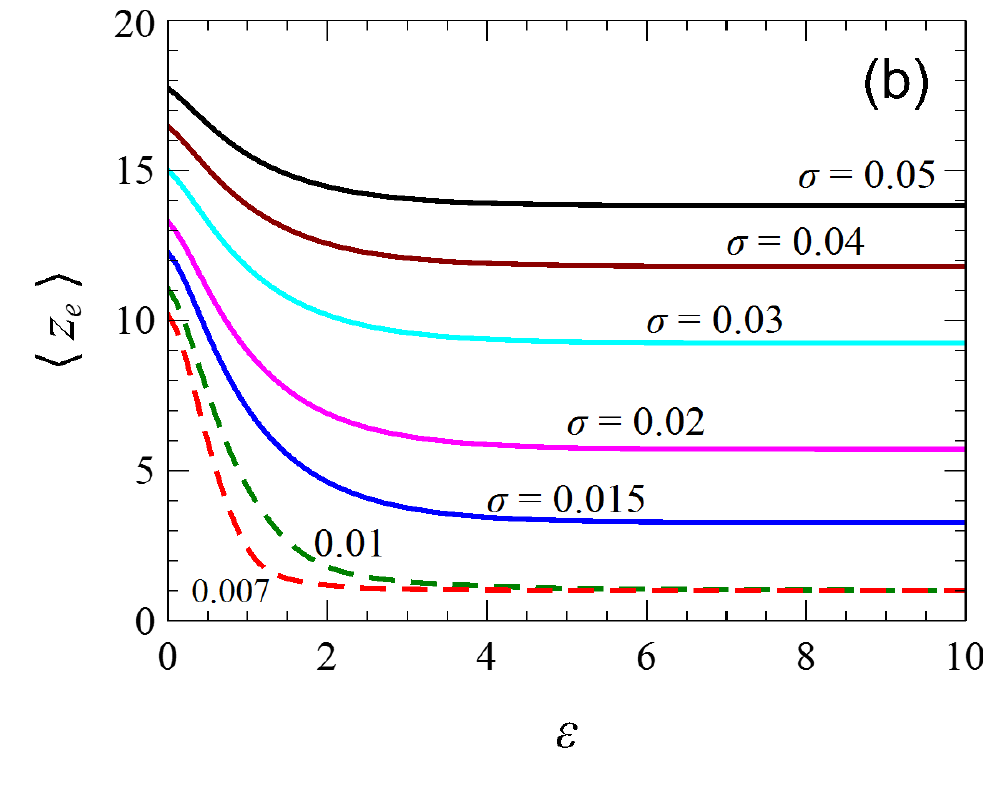}
\par\end{centering}
\caption{\label{fig:average height} Average distance of the free chain's ends
from the grafting surface in a monodisperse brush made of polymer
chains with $N=50$ (a) and $N=100$ (b) monomer units grafted at
the density $\sigma$ (indicated) as a function of the polymer-surface
adsorption energy $\varepsilon$. Solid and dashed curves correspond
to ``hairy'' and ``bald'' regimes, respectively.}
\end{figure}

As the monomer-substrate attraction strength changes, the brush undergoes
a certain restructuring. The brush thickness can be characterized
by the average height of the free ends, $\langle z_{e}\rangle$, which
decreases monotonically with an increase in the adsorption parameter,
$\varepsilon$, as demonstrated in Figure \ref{fig:average height}.
A saturation effect at large values, $\varepsilon\gtrsim3$, is due
to complete filling of the first layer by the adsorbed monomers. The
larger the area per chain, $1/\sigma$, the stronger the brush thickness
is affected by the attraction to the substrate. When the surface is
able to bind all the monomers, at sufficiently strong adsorption the
brush thickness reduces to one monomer length independently of the
chain length. Strictly speaking, these fully adsorbed states could
be laterally inhomogeneous if $1/\sigma$ is much larger than the
squared lateral size $R_{lat}^{2}\propto N^{3/2}$ of the effectively
2-dimensional adsorbed coil. However, the one-gradient version of
the SCF method does not allow to resolve lateral inhomogeneity and
is more suited to describe configurations when the $z-$profile of
the monomer density is formed by several overlapping chains.

\begin{figure}
\begin{centering}
\includegraphics[width=8cm]{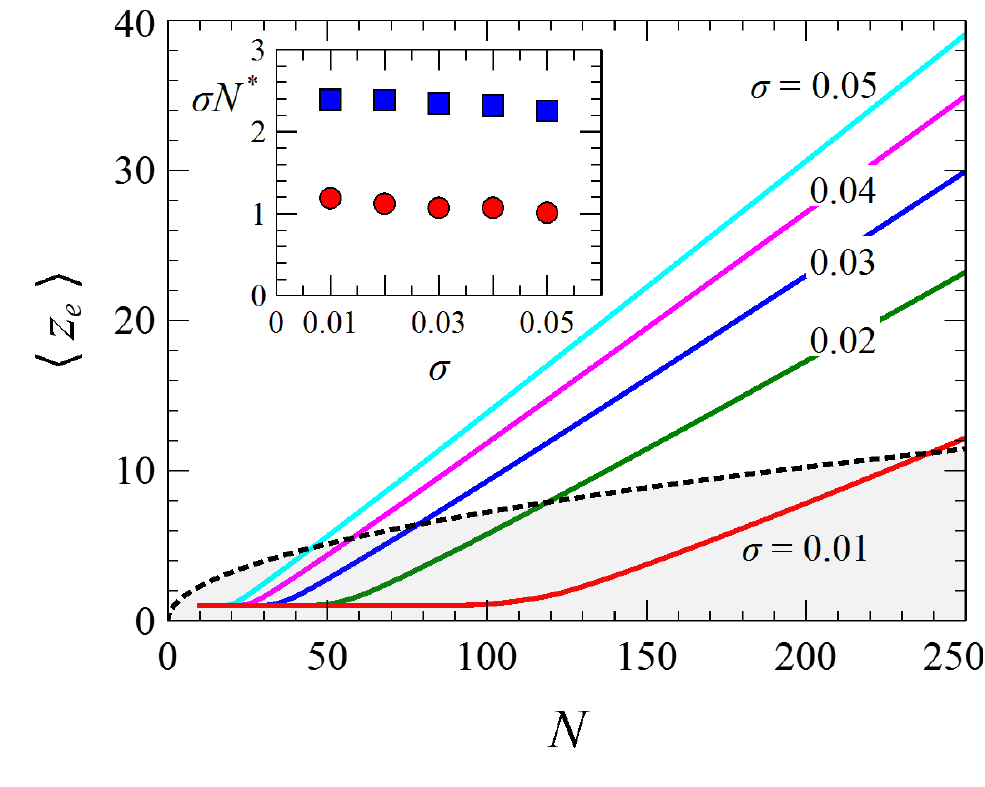}
\par\end{centering}
\caption{\label{fig:saturation height} Average distance of the free chain's
ends from the grafting surface in the saturation regime ($\varepsilon=5$)
as a function of the chain length, $N$, for several values of the
grafting density $\sigma$ (indicated in the figure). Black dashed
line shows the boundary $z=\sqrt{\pi N/6}$ between the mushroom and
brush regimes. Inset: the crossover values of the product $(\sigma N)^{*}\approx1$
separating the ``bald'', $\sigma N<1$ , and the ``hairy'' , $\sigma N>1$,
regimes (red circles) and mushroom and brush regimes (blue squares)
are both essentially independent of the brush grafting density, $\sigma$. }
\end{figure}

Figure \ref{fig:saturation height} displays the $N$- dependence
of the brush thickness in the saturation regime ($\varepsilon=5$)
for several values of the grafting density. Brushes composed of short
chains are completely adsorbed: no residual brush is left in the saturation
regime. Brushes composed of chains longer than a certain characteristic
value, $N^{*}(\sigma)$, cannot be fully adsorbed as the number of
monomers per unit area exceeds the adsorption capacity of the substrate.
Hence, even in the saturation limit, the surface retains non-adsorbed
chain tails, leading to an (approximately linear) increase in the
brush thickness with $N$. We will refer to these two saturation limits
as to the ``bald'' and ``hairy'' regimes. The inset shows that
the characteristic chain length separating the completely adsorbed
and the partially adsorbed regimes (red circles) satisfies the relation
$N^{*}\sigma\approx1$, so that the area per chain, $1/\sigma$, is
approximately equal to $N$. The condition $(\sigma N)^{*}=1$
marks a crossover point between regimes, which can be crossed both
by varying the chain length $N$ at fixed grafting density $\sigma$,
or by varying $\sigma$ at fixed $N$. 
Just from looking at Figs.\ \ref{fig:average height} and \ref{fig:saturation height},
the nature of the partially adsorbed state is not yet clear. Two scenarios
are conceivable: In the first one (Scenario 1), all chains are partly
adsorbed and contain $fN$ adsorbed monomers and desorbed tails of 
length $N(1-f)$. In this case, the desorbed tails (the hairs) form
an outer brush with effective grafting density $\sigma$ and effective
chain length $N_{eff}=N(1-f)$. In the second scenario (Scenario 2),
a fraction $q$ of chains is fully adsorbed, and the remaining ones
are fully desorbed and form the outer brush. The effective grafting
density of the outer brush is then reduced, $\sigmaeff=\sigma(1-q)$,
and the effective chain length is given by $N$. Below, we will show
that this second scenario applies in our system.

Due to the limitation of the SCF method, we also cannot
tell immediately whether the ``hairy'' state in the saturation
limit involves a well-formed residual brush or the non-adsorbed tails
form isolated ``mushrooms''. In order to address this question,
we compare the average free end height, $\left\langle z_{e}\right\rangle $
obtained from the SCF calculations with the analytical result $\left\langle z_{coil}\right\rangle =\sqrt{\pi N/6}$
for an isolated ideal coil of length $N$ grafted at the impenetrable
substrate. Hereafter, we use the criterion $\left\langle z_{e}\right\rangle =\left\langle z_{coil}\right\rangle $
to identify the boundary between the residual brush and mushroom regimes.
This boundary is shown in the inset of Figure \ref{fig:saturation height}
(blue squares) and is well described by the condition $(N\sigma)^{*}\approx2.5$.
In the rest of the paper we will focus on the regimes where
the residual brush exists at any adsorption strength, $N\sigma>2.5$.
A study of the other regimes at smaller grafting densities would require
a method allowing the resolution of laterally inhomogeneous structures.

\subsection{Self-consistent profiles of densities and fields} 

\begin{figure}
\begin{centering}
\includegraphics[width=7cm]{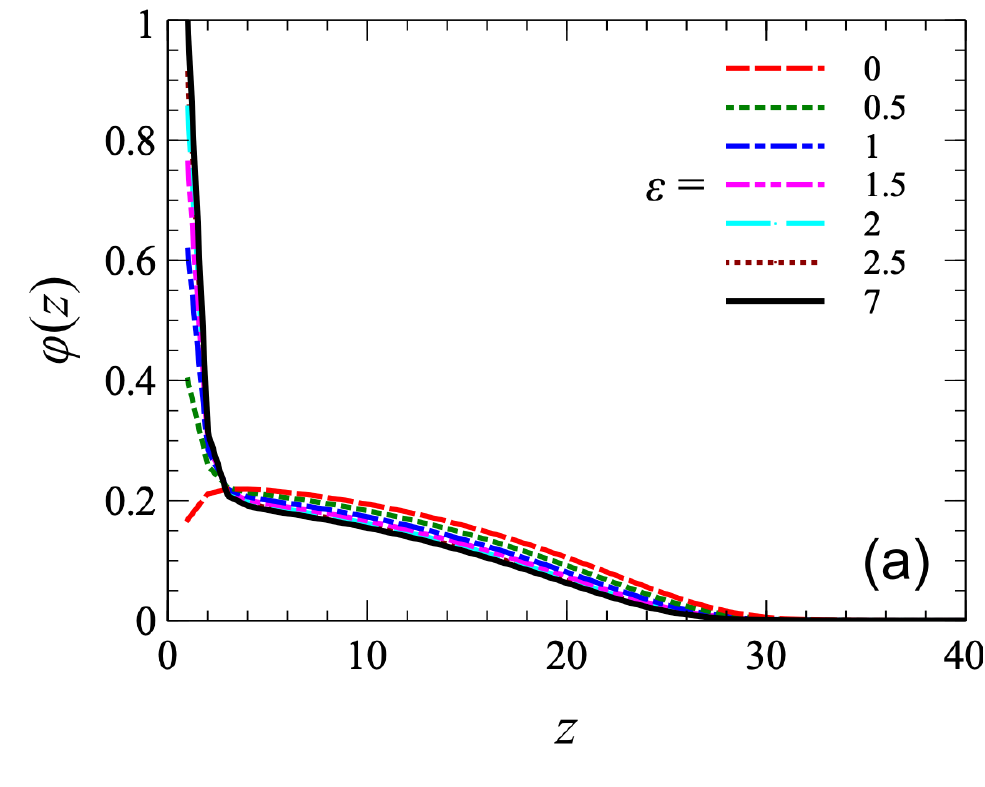} \includegraphics[width=7cm]{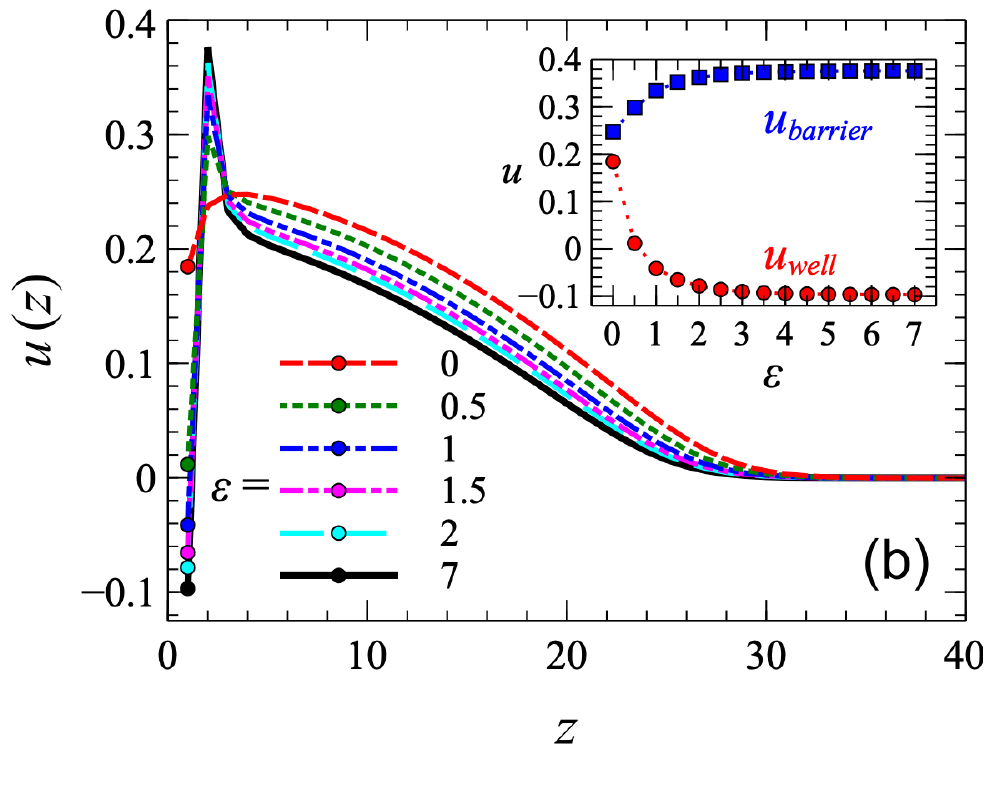}
\par\end{centering}
\caption{\label{fig:density profiles} Polymer density profiles, that
is, the volume fraction of segments $\varphi$ as a function of the
distance $z$ from the grafting surface (a) and the corresponding
self-consistent field potential profiles (b) for a monodisperse brush
made of polymer chains with $N=100$ monomer units grafted at the
density $\sigma=0.04$ at various values of the polymer-surface
adsorption energy $\varepsilon$, as indicated.  Inset to panel (b)
shows the values of the maximum (barrier) and the surface minimum
(well) as functions of the adsorption parameter
$\varepsilon\protect\geq0$. }
\end{figure}

The monomer density profiles change gradually with the increase
in the adsorption parameter, $\varepsilon$, as demonstrated in Figure
\ref{fig:density profiles}. A denser layer is formed in the nearest
vicinity of the grafting surface. The residual brush becomes thinner
in terms of its $z$-extension and simultaneously shows a decrease
in monomer density (except for the adsorbed layer). In an inert brush
with an intermediate density adequately described by the mean field
in the second virial approximation, the density profile coincides
with the potential of the mean force, $u(z)=v\varphi(z)$ where $v$
is the excluded volume parameter taken as $v$=1 in the present study.
The asymptotic shape of this field for well-formed brushes is given
by a well-known parabolic formula \citep{Milner:1988,Zhulina:1989}
$u(z)=\frac{3}{2}\left(\frac{\pi}{2}\sigma v\right)^{2/3}-\frac{3\pi^{2}}{8N^{2}}z^{2}$.
Note, however, that the analytical expression does not describe the
drop in the field in the closest vicinity to the substrate
which exists at $\varepsilon=0$ and is important in the context of
adsorption, see Figure \ref{fig:density profiles}b. With increasing
adsorption strength the density near the substrate increases and eventually
approaches dense packing. Clearly, the second virial approximation
is no longer valid and in the numerical SCF scheme, the Flory expression
relating the effective field to the local density, $u(z)=-\log\left[1-\varphi(z)\right]$,
is used. Importantly, the shape of the potential stabilizes at large
values of the adsorption parameter (in the saturation regime), and
the limiting shape is shown with $\varepsilon=7$ by the solid line.
Both the attractive well depth and the repulsive peak eventually stop
depending on $\varepsilon$ as shown in the inset.

\begin{figure}
\begin{centering}
\includegraphics[width=7cm]{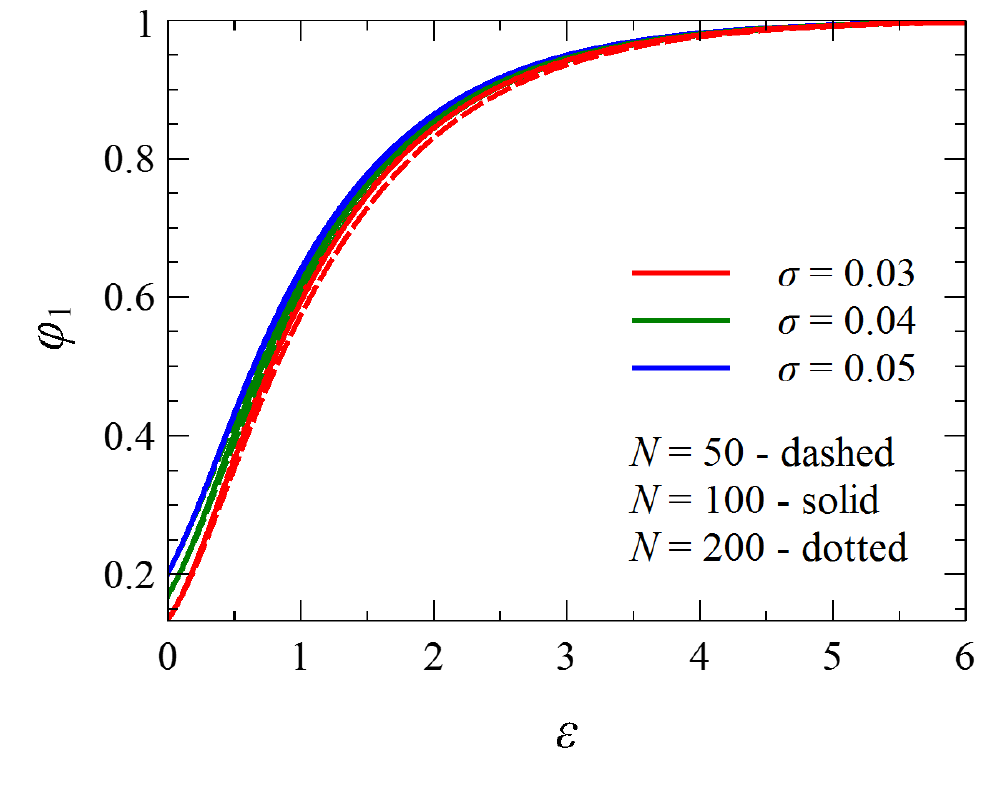} 
\par\end{centering}
\caption{\label{fig:first layer density} Polymer volume fraction at the surface,
that is, in the first adsorption layer $z=1$, in a monodisperse brush
made of polymer chains with $N=50$ (dashed lines), 100 (solid lines),
and 200 (dotted lines) monomer units grafted at the density $\sigma=0.03$
(red curves), 0.04 (green curves), and 0.05 (blue curves) as a function
of the polymer-surface adsorption energy.}
\end{figure}

Figure \ref{fig:first layer density} displays the monomer density
in the first adsorption layer, $\varphi_{1}$, as a function of the
adsorption parameter, $\varepsilon$, for different chain lengths
and grafting densities. The curves nearly collapse. The $N$-dependence
is negligible overall, and the dependence on $\sigma$ appears to
be noticeable only at small values of the adsorption parameter. Very
qualitatively, the curves resemble those of the Langmuir theory of
adsorption with its saturation effect and an approximately exponential
approach to the saturation value $\varphi_{1}=1$.

\subsection{Evidence for micro phase separation in the brush}

In order to obtain more insight into the adsorption scenarios,
we inspect the chain end distributions in the following Figure
\ref{fig:end distributions}. Figure \ref{fig:end distributions}a
displays the evolution of the chain end distributions with the increase
in the adsorption strength for the case of $N\sigma=4$. It is clear
that for all values of $\varepsilon\geq0.5$ the distribution is bimodal.
Figure \ref{fig:end distributions} suggests that one can identify
two coexisting phases, the adsorbed phase with the chain ends localized
near the substrate within a few monomers layers, and the brush phase
with a broad chain end distribution familiar for the neutral brushes.
The maximum describing the adsorbed chains becomes more prominent
with increasing $\varepsilon$. This is a clear indication
of the adsorption scenario 2 introduced in Section \ref{sec:regimes}:
At any given moment, a certain fraction of chains belongs to the
localized adsorbed phase with a large number of monomers in contact
with the surface, while the other chains form a brush with an effectively
reduced grafting density. If the fraction of adsorbed chains is $q,$
the effective grafting density of the residual brush is $\sigmaeff=\sigma(1-q)$.
In scenario 1, all chains would have similar conformations with
an adsorbed and desorbed chain block, and the chain end distribution
would always be unimodal. 

Since in a monodisperse brush all chains are identical, the actual
scenario means that chains can fluctuate between two states.
Figure \ref{fig:end distributions}b shows $-\ln P_{e}(z)$ which
can be interpreted as an effective potential landscape for single
chains. 
The two minima are separated by a very low barrier of no more than
1 $k_{B}T$ (see inset), suggesting that the states are not really
well separated and chains fluctuate rapidly between the two, without
being trapped in one state for long. Quite unexpectedly, this remains
true even when the adsorption parameter $\varepsilon$ is increased
up to 10 $k_{B}T$ or higher, since the shape of the distribution
stabilizes.\rev{On the other hand, this effect is in
line with de Gennes saturation argument for adsorbed polymer layers
whereby the adsorption free energy gain is outbalanced by steric repulsion~\cite{deGennes:1976, deGennes:1981, Bouchaud:1987, Descas:2006}.}

\begin{figure}
\begin{centering}
\includegraphics[width=8cm]{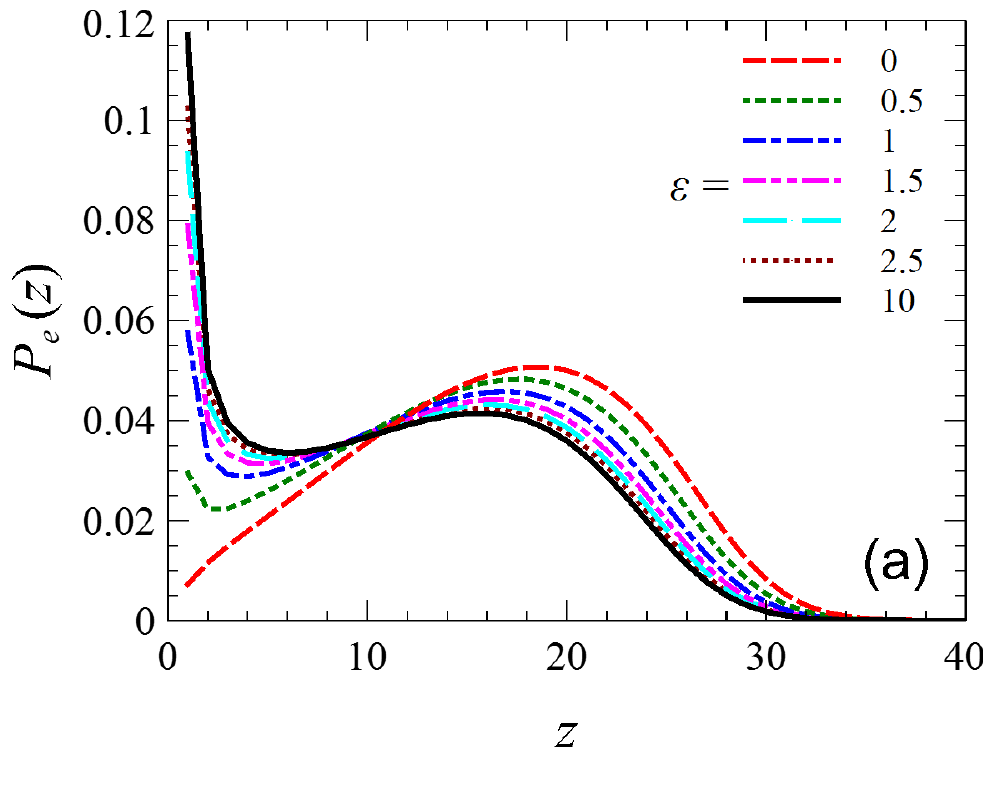} \includegraphics[width=8cm]{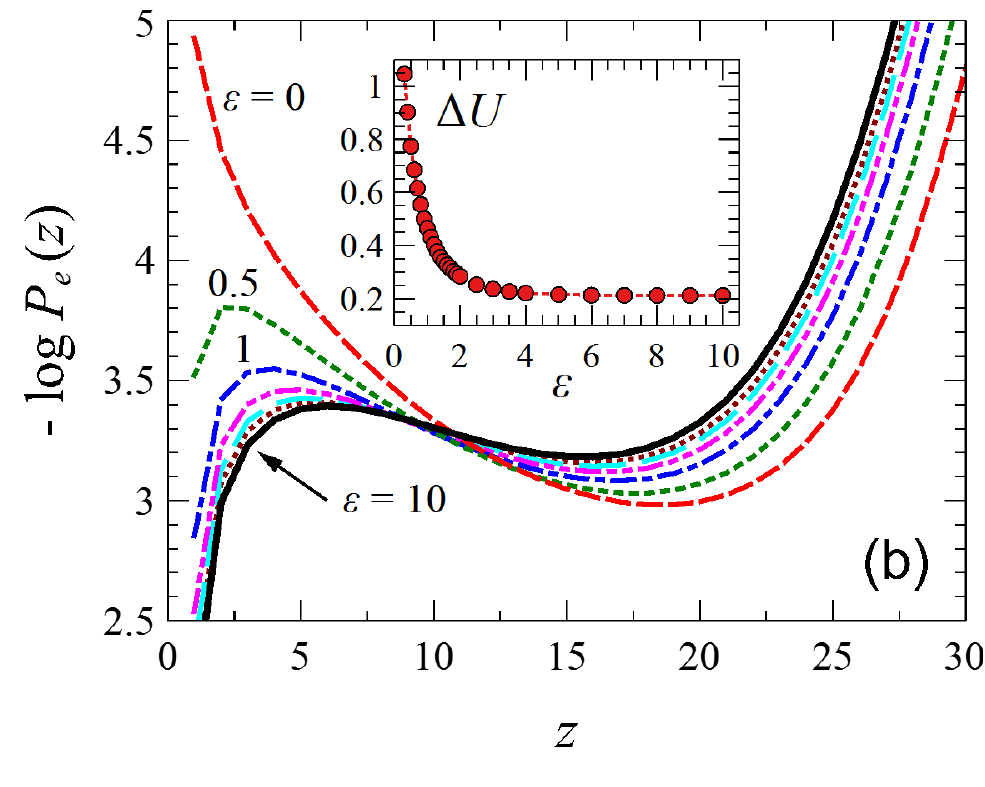}
\par\end{centering}
\caption{\label{fig:end distributions} Distributions $P_{e}(z)$ (a) and the
non-equilibrium free energy, $-\log P_{e}(z)$, (b) for the free ends
of brush chains in a monodisperse brush with $N=100$ and $\sigma=0.04$
at various polymer-surface adsorption energy $\varepsilon$, as indicated.
The inset shows the free energy barrier counted from the brush minimum.}
\end{figure}

\begin{figure}
\begin{centering}
\includegraphics[width=7cm]{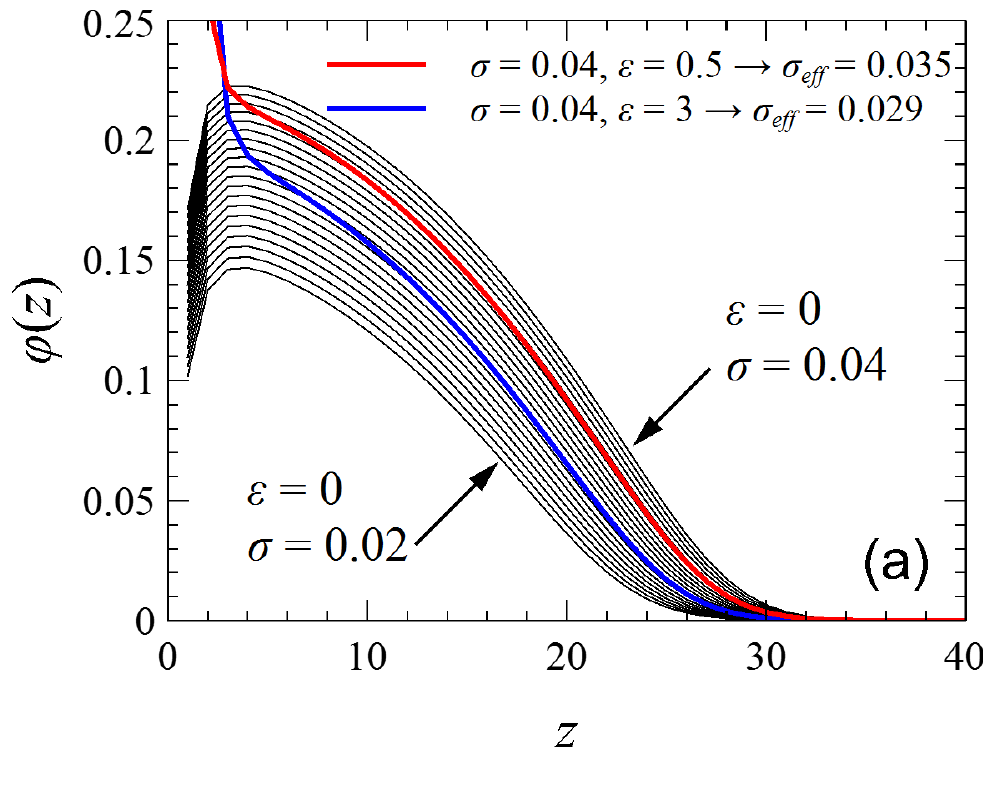} \includegraphics[width=7cm]{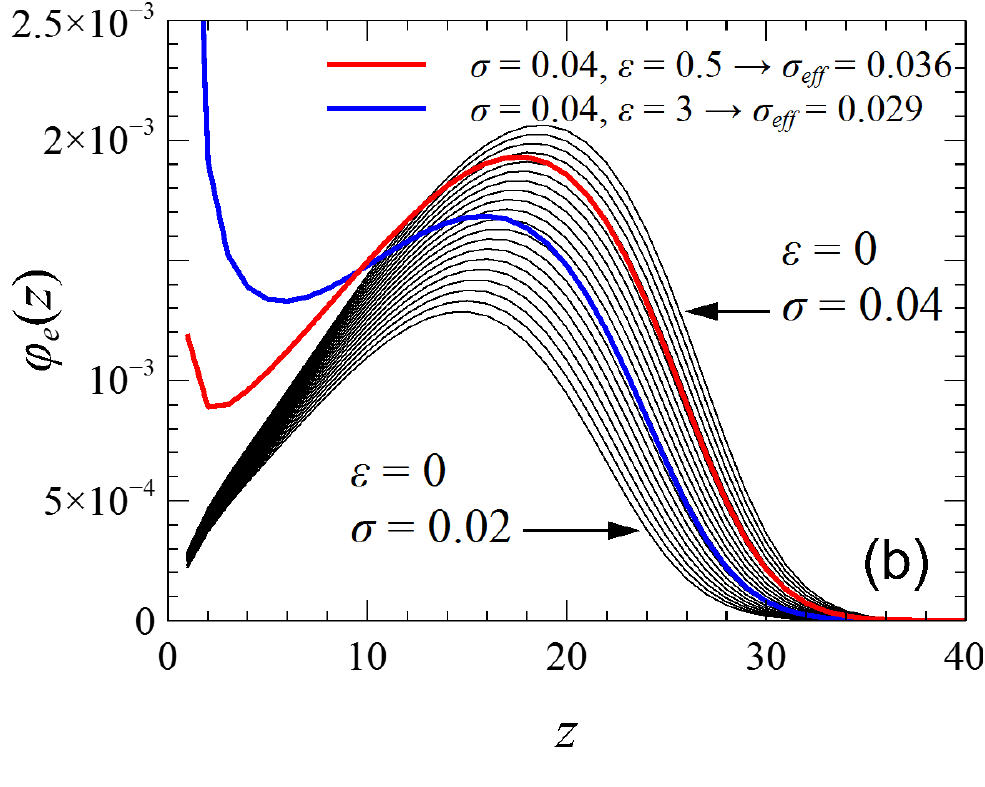}
\par\end{centering}
\caption{\label{fig:find_sigma_eff} Determination of the effective grafting
density of a polymer brush made of polymer chains with $N=100$ and
$\sigma=0.04$ at the adsorption energy $\varepsilon=0.5$ (red lines)
and 3 (blue lines) using overall polymer density profile (a) and free
ends' density profile (b). Thin black lines are the overall polymer
density profiles (a) and free ends' density profiles (b) in monodisperse
brushes with $N=100$ and neutral surface ($\varepsilon=0$) and grafting
density $\sigma=0.02$, 0,021, 0.022 ... 0.04. See explanation in
the text}
\end{figure}

For a monodisperse brush, the SF-SCF method does not allow to determine
the fraction of chains in the adsorbed phase, $q$, or $\sigmaeff$
directly. Nevertheless, we can suggest a way of determining $q$ or
$\sigmaeff$ based on the similarity between the residual brush and
a conventional brush grafted onto a non-attractive surface. The 
procedure is illustrated by Figure \ref{fig:find_sigma_eff}. Here 
we show the full density profiles $\varphi\left(z\right)$ (panel a) 
and the density profiles of free ends $\varphi_{e}\left(z\right)$
(panel b) 
for brushes with $\varepsilon=0.5$ (red curves) and $\varepsilon=3$
(blue curves) grafted at $\sigma=0.04$ plotted together with those
for the \rev{non-adsorptive (neutral)} brush ($\varepsilon=0$)
in the range of $\sigma=0.02$ ... 0.04 with $\Delta\sigma=0.001$
for the sake of better visibility; in the actual fitting procedure
we used a 10 times smaller step $\Delta\sigma=0.0001$.

To extract $\sigmaeff$, we compare the density profile
and ends distributions in a brush grafted the density $\sigma$ at
$\varepsilon>0$ with the profiles obtained for the brush at $\varepsilon=0$
and various grafting densities $\sigma$(Figure \ref{fig:find_sigma_eff}a).
We calculate the mean-square distance between two profiles defined
as

\begin{equation}
(\Delta\varphi)^{2}(\sigma_{1})=\sum_{z\geq z_{0}}[\varphi(z,\sigma,\varepsilon)-\varphi(z,\sigma_{1},0)]^{2}\label{eq:msq distance}
\end{equation}
for various values of $\sigma_{1}$ and identify $\sigmaeff$ as the
value that minimizes the mean-square distance $(\Delta\varphi)^{2}$.
Here, $z_{0}$ is chosen to be the maximum of the end density
profile (Fig.\ \ref{fig:find_sigma_eff} b), thus the mean-square
distance was only determined in the range of $z$-values where the
neutral brush and the partially adsorbed brush profiles where directly
comparable. This procedure was applied both to the full
density profiles $\varphi(z)$ and the density profiles of the free
ends $\varphi_{e}\left(z\right)$ (Figure \ref{fig:find_sigma_eff}
a and b).


\begin{figure}
\begin{centering}
\includegraphics[width=7cm]{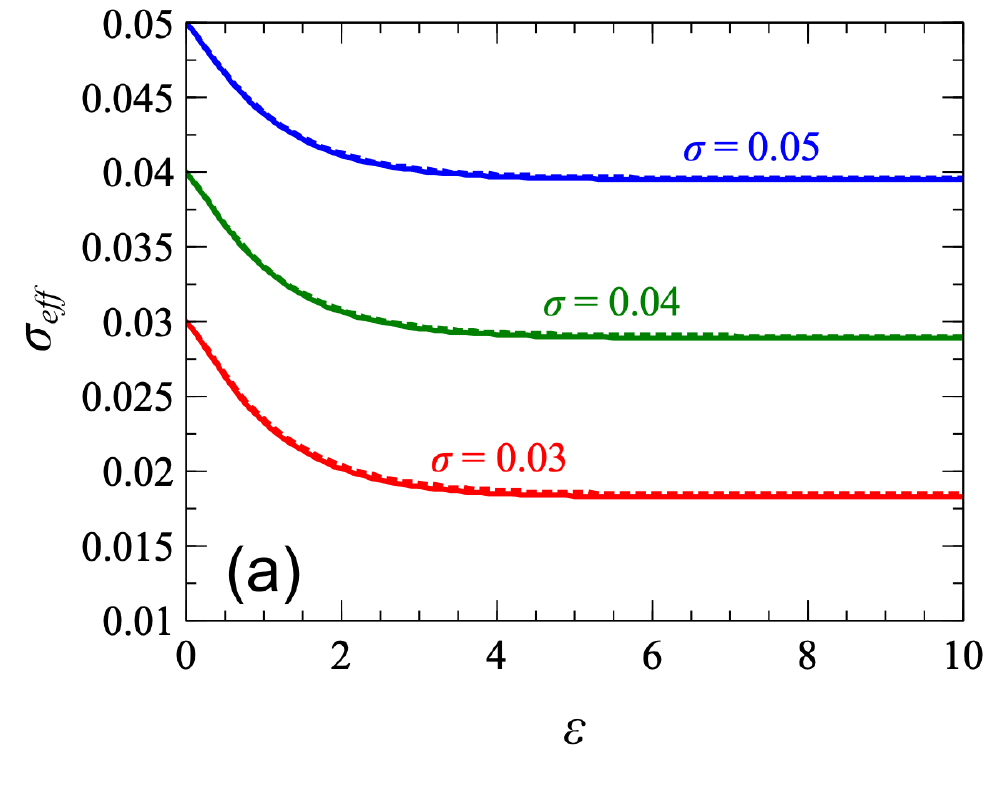} \includegraphics[width=7cm]{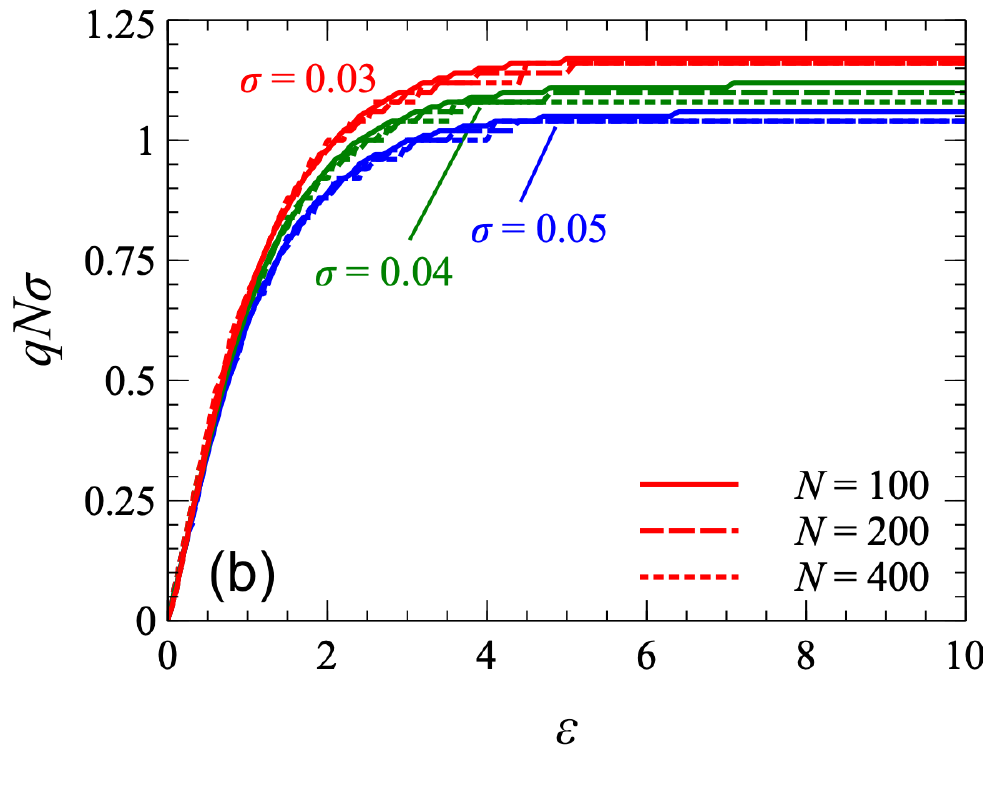}
\par\end{centering}
\caption{\label{fig:effective sigma}
(a) Effective grafting density $\sigmaeff$ in a brush made of
polymers with chain length $N=100$ and different grafting densities
$\sigma$ as indicated. Here $\sigmaeff$ is determined by fitting full
monomer density profiles (solid lines) and end density profiles
(dashed lines) as described in the main text.  (b) Rescaled fraction
$q$ of chains in the adorbed phase as a function of $\varepsilon$ for
different chains lengths $N$ and grafting densities $\sigma$ as
indicated.
}
\end{figure}

The dependence of the effective grafting density $\sigmaeff$ determined
in this way on the adsorption energy $\varepsilon$ is shown in Figure
\ref{fig:effective sigma}a. 
The curves were obtained by fitting the monomer density profiles (solid
lines) and the end distributions (dotted lines) and one can see an
excellent agreement between these two approaches. With increasing
adsorption energy, $\sigmaeff$ monotonically decreases and reaches
saturation. In this case, the range of decrease in effective density,
$\text{\ensuremath{\sigma-}\ensuremath{\sigmaeff} }$, is roughly
determined by the inverse chain length $1/N$. A basic explanation
is provided by the following argument. If we assume that in the saturation
regime a chain in the adsorbed phase is completely adsorbed on the
surface, then it occupies $N$ lattice cells in the first lattice
layer, or the surface area is equal to $N$. The number of
grafted chains per this surface area is $N\sigma$. Hence, the fraction
of chains in the adsorbed phase in the saturation regime, $q=1-\sigmaeff/\sigma$,
is equal to $q^{*}=1/(N\sigma)$, and the effective grafting density
is thus $\sigmaeff=\sigma(1-q^{*})=\sigma-1/N$. To test this
argument, we show in Figure \ref{fig:effective sigma}b the rescaled
fraction of adsorbed chains, $q\sigma N$, as a function of adsorption
strength $\varepsilon$ for various chain lengths $N$ and grafting
densities $\sigma$. As expected, it initially grows with $\varepsilon$
and then saturates at a value close to 1. However, the saturation
value is found to exceed 1, especially for smaller $\sigma$.  
This would imply that the fraction of chain contacts with the surface
in the adsorbed phase is smaller than unity.

The average fraction of adsorbed units per chain \emph{in the brush as
a whole} is given by $\langle\theta\rangle=\varphi_{1}/(N\sigma)$
where $\varphi_{1}=\varphi(1,\sigma,\varepsilon)$ is the monomer
density in the first layer $z=1$. To obtain the fraction of adsorbed
units for a chain \emph{in the adsorbed phase}, $\thetaads$, we should
take into account only the fraction of chains in the adsorbed phase
$q$ and eliminate the contribution of monomers that belong to chains
in the residual brush but are still in contact with the substrate.
Then the fraction of monomer-surface contacts in an adsorbed chain is
given by 
\begin{equation}
\thetaads=\frac{\varphi(1,\sigma,\varepsilon)-\varphi(1,\sigmaeff,0)}{N\sigma q}.\label{eq:theta}
\end{equation}

\begin{figure}
\begin{centering}
\includegraphics[width=7cm]{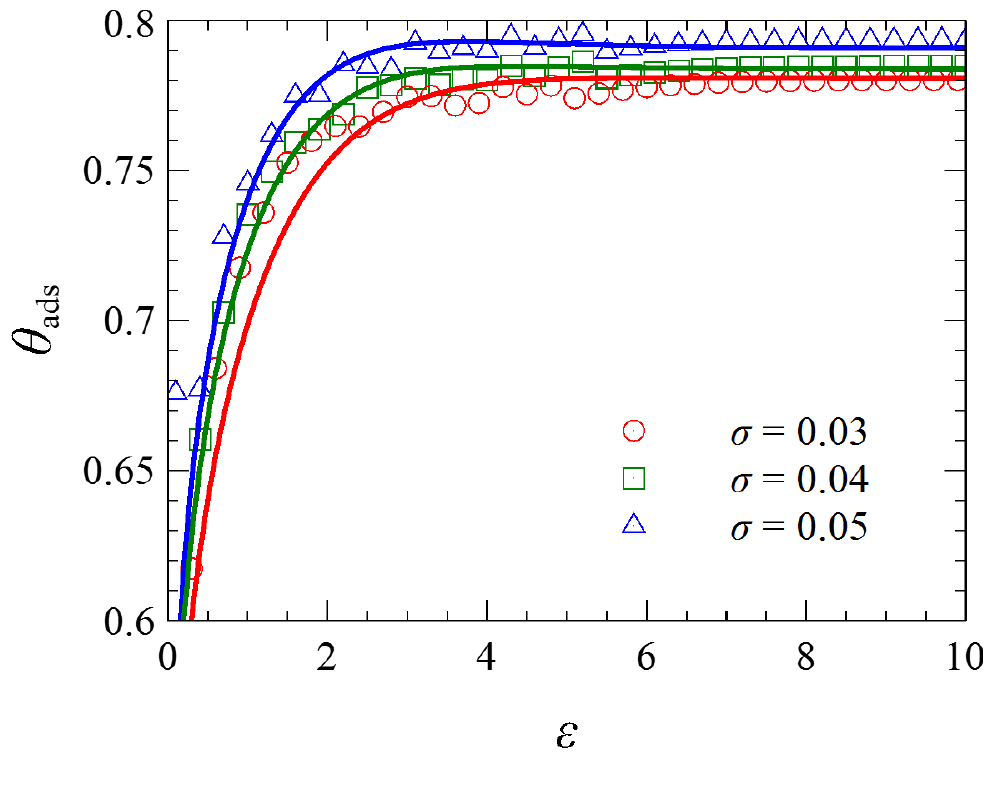}
\par\end{centering}
\caption{\label{fig:theta in ads phase}Fraction of contacts in a chain belonging
to the adsorbed phase, $\thetaads$, as a function of the adsorption
parameter $\varepsilon$ in a monodisperse polymer
brush with $N=100$ grafted at the density $\sigma$ as indicated.
Symbols show the result of direct calculation of $\thetaads$ according
to Eq. \ref{eq:theta} with the values of $\varphi(1,\sigma,\varepsilon)$,
$\varphi(1,\sigmaeff,0)$, and $q$ obtained by the minimization
procedure described in the text. Solid curves are obtained with the
help of phenomenological fitting for $q(\varepsilon)$ and
$\varphi(1,\sigma,\varepsilon)-\varphi(1,\sigmaeff,0)$.}
\end{figure}

Figure \ref{fig:theta in ads phase} shows the dependence of $\thetaads$on
the adsorption energy $\varepsilon$ calculated according to Eq. (\ref{eq:theta})
using the data presented in . \ref{fig:first layer density}
and \ref{fig:effective sigma}. The symbols are the result
of direct application of Eq. (\ref{eq:theta}) to the data points
for $\varphi(z=1,\sigma)$, $\varphi_{\varepsilon=0}(1,\sigmaeff)$,
and $q$. We have also approximated $q(\varepsilon)$ and $\varphi(1,\sigma,\varepsilon)-\varphi(1,\sigmaeff,0)$
by simple phenomenological functions to obtain the curves shown in
Figure \ref{fig:theta in ads phase} by solid lines.

The dependence of the fraction of contacts in adsorbed chains
$\thetaads$ on the adsorption energy $\varepsilon$ at various grafting
densities $\sigma$ shown in Figure \ref{fig:theta in ads phase}
demonstrate that $\thetaads$ increases with increasing $\varepsilon$
reaching a plateau value, which is less than one. This indicates that
the chains in the adsorbed phase are not fully adsorbed, even in the
saturation regime, and have a small fraction of non-adsorbed units
forming loops and tails. The fraction of contacts in adsorbed chains
also weakly increases with increasing grafting density. It is also
worth noting that already at rather \rev{small
values $\varepsilon=0.2-0.3$ the fraction} of contacts in the adsorbed
chains is high: $\thetaads\approx0.6$ .

\section{Scaling analysis}

\label{sec:scaling_regimes}

In order to obtain more physical insight in the nature of the
partially adsorbed state, we will now present a scaling analysis of
the system, where we combine blob concepts for adsorbing chains and of
polymer brushes \cite{Halperin:1994,Rubinstein_book,Descas:2004}.  As
before, we consider a monodisperse brush made of chains of length $N$
in good solvent, grafted with density $\sigma$ on an adsorbing
substrate with adsorption strength $\varepsilon$. Depending on the
chain length and the grafting density we distinguish between different
regimes as described in Sec.\ \ref{sec:regimes}. We will also discuss
the two possible scenarios of partial adsorption introduced in Sec.\
\ref{sec:regimes}.

\begin{figure}
\begin{centering}
\includegraphics[width=7cm]{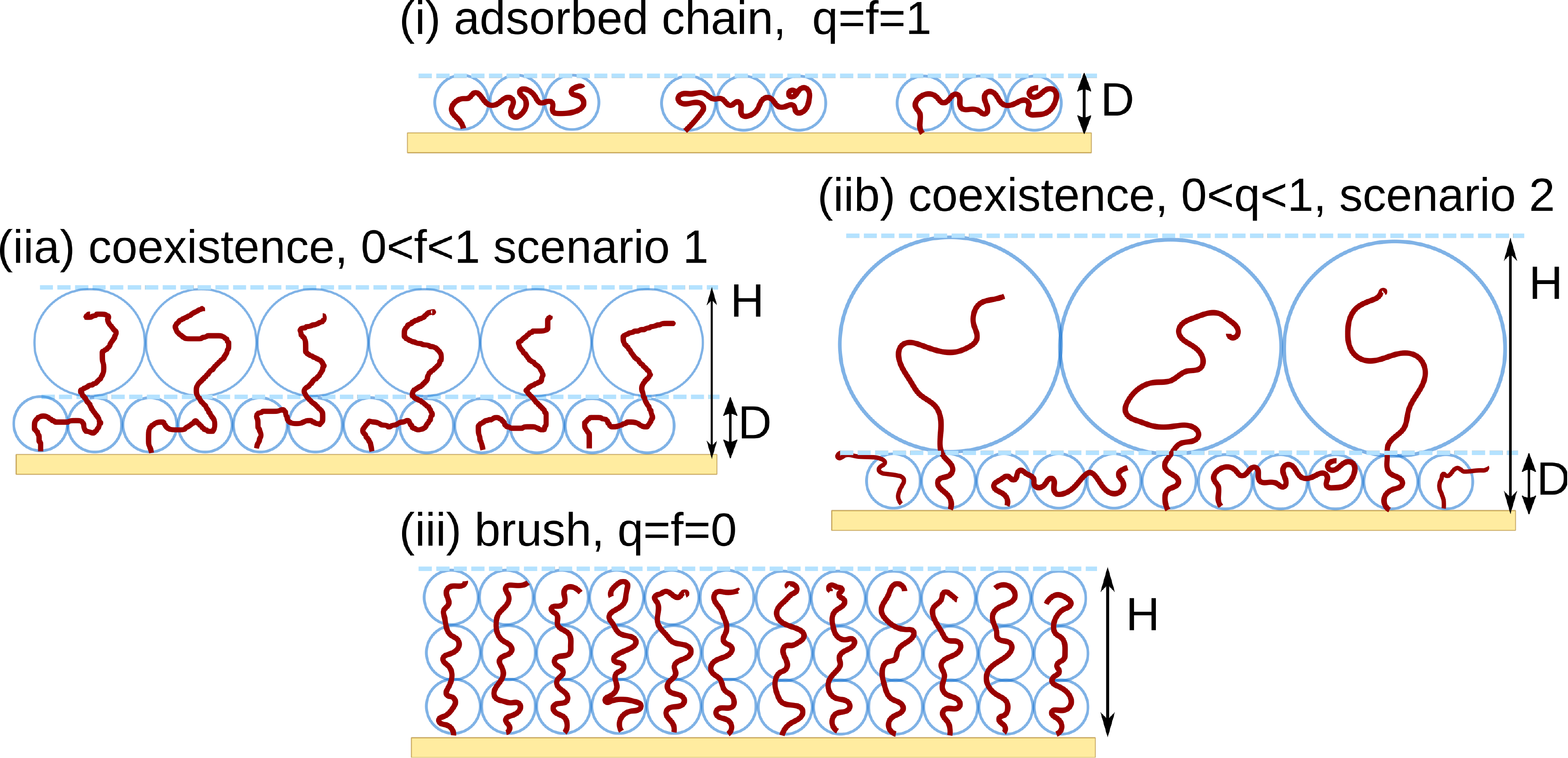}
\par\end{centering}
\caption{\label{fig:cartoon}
Cartoon showing different regimes in the adsorbing brush. See text
for explanation.}
\end{figure}

\subsection{Scaling regimes}

\subsubsection{Regime (i) -- Low grafting densities}

At \rev{very} low grafting densities, the grafted chains do not
interact with each other and they behave like isolated adsorbing
chains.

We begin with briefly recapitulating the simplest scaling picture for
a single adsorbed chain \cite{Rubinstein_book}. The chain is pictured
as a chain of blobs lying on the surface. Every blob has the size $D$
(in units of $a$) and contains $m\sim D^{1/\nu}$ monomers, where $\nu
\approx 3/5$ is the Flory exponent. Thus, the total chain has $N/m$
blobs. According to the blob picture, every blob carries an entropic
free energy penalty $k_{B}T$ to the free energy, resulting in a total
entropic contribution $F_{e}\sim N/m\sim ND^{-1/\nu}$ (in units of
$k_{B}T$). On the other hand, the fraction of monomers in contact with
the surface is estimated as \rev{$D^{(\phi-1)/\nu}$
\cite{deGennes:1983b,Descas:2006}} and the energy gain due to
adsorption is hence given by \rev{ $F_{a}\sim-\varepsilon N
\:D^{(\phi-1)/\nu}$, where $\phi$ is the crossover exponent. The value
of $\phi$ is still debated in the literature. Below, we will
approximate it with $\phi \approx 1/2$ which is close to the values
obtained from numerical simulations \cite{Grassberger:2005,Klushin:2013,Zhang:2018}.}
Taking everything together, the total energy is estimated as 
\begin{equation}
F\sim N\:D^{-1/\nu}+\varepsilon N\:\rev{D^{(\phi-1)/\nu}}
\end{equation}
Minimizing this expression with respect to $D$, we obtain 
\begin{equation}
D\sim \rev{\varepsilon^{-{\nu}/{\phi}}}
  \sim \rev{\varepsilon^{-6/5}}
\label{eq:adsorption_blob}
\end{equation}
The chain can be considered adsorbed if it contains several adsorption
blobs ($N/m\gg1$), which implies 
\begin{equation}
   N\:D^{-1/\nu} \sim 
     N\:\rev{\varepsilon^{1/\phi}} \sim N\: \rev{\varepsilon^{2}}
      \gg1
  \label{eq:ads_condition}
\end{equation}
The total area covered by the chain on the substrate
is estimated as 
\begin{equation}
  A_{c}\sim N/m\:D^{2} \sim 
   N\: \rev{\varepsilon^{{(1-2\nu)}/{\phi}}}
   \sim N\: \rev{\varepsilon^{-2/5}}
   \label{eq:Ac}
\end{equation}

\rev{As the grafting density $\sigma$ increases, the  adsorbed chains
start to interact and the two dimensional blob chain conformations on
the surface influence each other. A detailed scaling analysis of this
regime has been carried out by Descas {\em et al}~\cite{Descas:2006}. 
Eventually, the surface is fully covered by adsorption blobs and 
saturates. The crossover to the oversaturated regime is reached at
$\sigma \: A_c \sim 1$ (where $A_c$ is given by Eq.\ (\ref{eq:Ac})),
hence \cite{Descas:2006}}
\begin{equation}
 (\sigma N)^{\star} 
   \sim \rev{\varepsilon^{{(2\nu-1)}/{\phi}}}
   \sim \rev{\varepsilon^{2/5}}
   \label{eq:crossover}
\end{equation}

\subsubsection{Regime (ii) -- Intermediate grafting densities}

Once the overlap concentration is reached, one of the two possible
scenarios of partial adsorption discussed in Sec. \ref{sec:regimes}
applies. We will discuss them one after the other.

In Scenario 1 (iia in Fig.\ \ref{fig:cartoon}), a fraction $f$
of monomers in each chain adsorbs to the substrate. The remaining
tails comprising $N(1-f)$ monomers desorb and together form an outer
brush. In the blob picture, the substrate is thus covered by a dense
layer of small adsorbed blobs of size $D$ containing $m$ monomers
each, covered by a more tenuous layer of larger blobs of size $\sigma^{-1/2}$.
The fraction $f$ of adsorbed monomers is estimated as follows: The
number of surface blobs per area is given by $(1/D)^{2}=\sigma fN/m$.
Inserting $m\sim D^{1/\nu}$, one obtains 
\begin{equation}
D\sim(\sigma fN)^{-\frac{\nu}{2\nu-1}}\sim(\sigma fN)^{-3}\label{eq:D_f}
\end{equation}
and the adsorption energy per chain 
$F_{a}\sim-\varepsilon\sigma fN \rev{D^{(\phi-1)/\nu}}
\sim-\varepsilon(\sigma fN)\rev{{}^{\frac{2\nu-\phi}{2\nu-1}}}$.
The entropy cost per area is given by the total number of blobs per
area, which includes surface (adsorption) blobs and outer brush blobs.
The number of adsorption blobs is given by $(1/D)^{2}\sim(\sigma fN)^{\frac{2\nu}{2\nu-1}}$.
The number of brush blobs in the outer layer corresponds to that of
an Alexander brush with grafting density $\sigma$ and effective chain
length $N(1-f)$ and is hence given by\cite{Halperin:1994} $N(1-f)\sigma^{\frac{2\nu+1}{2\nu}}$.
Taking everything together, one gets the free energy 
\begin{equation}
F\sim-\varepsilon(\sigma fN)\rev{{}^{\frac{2\nu-\phi}{2\nu-1}}}
  +(\sigma fN)^{\frac{2\nu}{2\nu-1}}
  +N(1-f)\sigma^{\frac{2\nu+1}{2\nu}}\label{eq:energy_scen1}
\end{equation}
In the limit $N\to\infty$, the last term becomes negligible compared
to the first two terms, and it suffices to minimize 
$F\sim-\varepsilon(\sigma fN)\rev{{}^{\frac{2\nu-\phi}{2\nu-1}}}
  +(\sigma fN)^{\frac{2\nu}{2\nu-1}},$
which results in 
\begin{equation}
(\sigma fN)\sim \rev{\varepsilon^{{(2\nu-1)}{\phi}}}
  \sim(\sigma N)^{*},\label{eq:f}
\end{equation}
where $(\sigma N)^{*}$ is the crossover parameter (\ref{eq:crossover})
that separates regime (i) and (ii).

In Scenario 2 (ii b in Fig.\ \ref{fig:cartoon}), a fraction $q$
of chains remain fully adsorbed, whereas the remaining chains desorb
fully and have no contacts to the surface. The resulting picture is
similar to Scenario 1 in many \rev{respects}. The surface
is still first covered by a dense layer of adsorbed small blobs (thickness
$D$), followed by a dilute layer of large blobs, which now have the
size $[(1-q)\sigma]^{-1/2}$. The number of surface blobs per area
is now given by $(1/D)^{2}=\sigma qN/m+\sigma(1-q)$. In the limit
of large $N$, the second term can be neglected. Using $m\sim D^{1/\nu}$,
we obtain 
\begin{equation}
D\sim(\sigma qN)^{-\frac{\nu}{2\nu-1}}\sim(\sigma qN)^{-3},\label{eq:D_q}
\end{equation}
which has the same form than Eq.\ (\ref{eq:D_f}) with $f$ replaced
by $q$. Accordingly, the adsorption energy per chain is 
$F_{a}\sim-\varepsilon(\sigma qN)\rev{{}^{\frac{2\nu-\phi}{2\nu-1}}}$
and the number of adsorption blobs is $(1/D)^{2}\sim(\sigma qN)^{\frac{2\nu}{2\nu-1}}$.
The number of blobs in the outer layer is that of a regular brush
with effective grafting density $\sigma q$ and chain length $N$,
resulting in $N[(1-q)\sigma]^{\frac{2\nu+1}{2\nu}}$. Thus the total
free energy is estimated as 
\begin{equation}
  F\sim-\varepsilon(\sigma qN)\rev{{}^{\frac{2\nu-\phi}{2\nu-1}}}
    +(\sigma qN)^{\frac{2\nu}{2\nu-1}}
    +N[(1-q)\sigma]^{\frac{2\nu+1}{2\nu}},
    \label{eq:energy_scen2}
\end{equation}
which has the same form than Eq.\ (\ref{eq:energy_scen1}) with $f$
replaced by $q$, except for the last term. However, the last term
again vanishes in the limit $N\to\infty$ in relation to the other
two, and can be neglected. Thus the remaining calculation is the same
as in scenario 1 (iia in Fig.\ \ref{fig:cartoon}), and one obtains 
\begin{equation}
(\sigma qN)\sim\varepsilon\rev{{}^{{(2\nu-1)}/{\phi}}}
  \sim(\sigma N)^{*}.\label{eq:q}
\end{equation}
Here, $(\sigma N)^{*}$ is again the parameter characterizing
the crossover from regime (i) and (ii) (Eq.\ (\ref{eq:crossover})).

Comparing the final free energies in Scenario 1 and Scenario 2, we
find that they only differ in the free energy contribution of the
outer brush, which is given by $N(1-f)\sigma^{\frac{2\nu+1}{2\nu}}$
in Scenario 1 and by $N[\sigma(1-q)]^{\frac{2\nu+1}{2\nu}}$ in Scenario
2. According to Eqs.\ (\ref{eq:f}) and (\ref{eq:q}), we have $f=q$.
From $0<(1-f)=(1-q)<1$, we get $(1-q)^{\frac{2\nu+1}{2\nu}}<(1-f)$,
hence Scenario 2 is predicted to be more favorable than Scenario 1
in agreement with the SCF results.

An important result from the scaling analysis is that the thickness
of the surface layer is independent of $\sigma$ (in both scenarios).
The grafting density of chains in the adsorbed state remains constant
and corresponds to the crossover grafting density, $\sigma q\sim\sigma^{\star}$.
Setting $q=1$ in Eq.\ (\ref{eq:q}), one recovers the expression
Eq.\ (\ref{eq:crossover}) for the crossover grafting density $\sigma^{\star}$.

\subsubsection{Regime (iii) -- High grafting densities}

At high grafting densities, all chains are fully desorbed. The transition
to this state takes place at the grafting density where the blob size
of a pure brush equals the size of the adsorption blob, 
$\sigma^{-1/2}=D\sim\varepsilon\rev{{}^{-{\nu}/\phi}}$.
This second crossover grafting density is thus given by 
\begin{equation}
\sigma_{0}\sim \rev{\varepsilon^{{2\nu}/\phi}}
  \sim \rev{\varepsilon^{12/5}}
\end{equation}
As a consistency check, we can insert this value in the expression for
$q$ in Eq.\ (\ref{eq:q}) and obtain
$q\sim1/(N \rev{\varepsilon^{1/\phi}})\ll1$ where the last inequality
results from Eq.\ (\ref{eq:ads_condition}).  Hence $q\approx 0$ at the
transition as it should.  \rev{Note that the scaling approach is
limited to smaller values of $\varepsilon$ when a single adsorption
blob comprizes at least a few monomers.}

\subsection{Comparison with SCF results}

Based on the results from the previous section, we can now compare
the predictions of the scaling analysis with the results from the
SCF calculations.

First, we remark that the full series of transitions from {\em (i)
adsorbed} via {\em (ii) partially adsorbed} to {\em (iii) desorbed}
discussed above can only be observed if $\sigma$ or $\varepsilon$
are varied, but not if $N$ is varied at fixed $\sigma$ and $\varepsilon$.
In the latter case, the phase behavior depends on the value of 
\rev{$x:=\varepsilon\sigma^{-\phi/2 \nu} \sim \varepsilon
\sigma^{-5/12}$}.
If $x\ll1$ (weak adsorption), the chains never adsorb, and the grafted
polymer layer undergoes a regular transition from a mushroom to a
brush with increasing $N$. If $x\gg1$ (strong adsorption), the surface
is always covered by an adsorbed layer. Upon increasing $N$ in that
case, one expects a transition from (i) (fully adsorbed layer), to
(ii) (partially adsorbed layer), but the pure brush state is never
reached. This situation is studied in Fig.\ \ref{fig:saturation height}.
The height of the brush as a function of chain length is expected
to behave as 
\begin{equation}
H\sim\left\{ \begin{array}{ll}
D\sim \rev{\varepsilon^{-6/5}} & :\;
  N<N^{\star}\sim \rev{\varepsilon^{2/5}}/\sigma\\
N[\sigma-\sigma^{\star}]^{1/3} & :\;N>N^{\star}
\end{array}\right.,
\end{equation}
which explains the crossover from constant to linear behavior as a
function of $N$ observed in Fig.\ \ref{fig:saturation height}.
The scaling theory predicts the transition point to be independent
of $\sigma$, $(\sigma N)^{*}\sim \rev{\varepsilon^{2/5}}$, in agreement
with the SCF results.

When varying $\varepsilon$ at fixed $\sigma$ and $N$, the scaling
theory predicts the regime of partial adsorption to be very broad in
the limit of large $N$, ranging from $\varepsilon_{0}\sim
\rev{\sigma^{5/12}}$ to $\varepsilon^{*}\sim (\sigma
N)\rev{{}^{5/2}}$. This is confirmed by the SCF results, e.g., Figs.\
\ref{fig:first layer density}, \ref{fig:effective sigma}, and
\ref{fig:theta in ads phase}, where the transition to the pure brush
cannot be clearly localized.

The fraction of adsorbed chains is predicted to scale as
$q\sim\frac{1}{\sigma N}\rev{\varepsilon^{2/5}}$, which is consistent
with the SCF data in Fig.\ \ref{fig:effective sigma}.  In particular,
the data for $q\sigma N$ for different $\sigma$ and $N$ as a function
of $\varepsilon$ roughly collapse as predicted (see Fig.\
\ref{fig:effective sigma} b)).

In other respect, the comparison between the SCF results and the
scaling theory is only partly convincing. For example, the density of
adsorbed contacts $N\sigma q \rev{D^{(\phi -1)/\nu}} \sim
\rev{\varepsilon^{7/5}}$ is predicted to be independent of $N$ and
$\sigma$, which is in agreement with the SCF results. However, the
predicted dependence $\sim \rev{\varepsilon^{7/5}}$ is not reflected
in the data of Fig.\ \ref{fig:first layer density}.  Likewise, the
fraction of contacts in the adsorbed phase is predicted to be
$\thetaads\sim \rev{D^{(\phi-1)/\nu} \sim \varepsilon^{(1-\phi)/\phi}
\sim \varepsilon}$, independent of $\sigma$.  The corresponding SCF
data also seem to roughly collapse for different $\sigma$ according to
Fig. \ref{fig:theta in ads phase}. However, the data do not reflect
the scaling law $\sim \rev{\varepsilon}$.

Thus the results of the scaling analysis do qualitatively agree with
the SCF results, however, the actual dependence of various quantities
on $\varepsilon$ is poorly captured. One likely explanation is that
the fraction of adsorbed monomers and the size of the adsorption blob
saturate at high $\varepsilon$. Indeed, looking at Fig.\
\ref{fig:density profiles}, one gets the impression that the thickness
of the adsorbed layer approaches a constant for $\varepsilon>0.5$
and that this constant is of the order of the lattice constant $a$.

We can take such saturation effects into account by assuming that the
thickness of the adsorbed layer is a general function
$D(\varepsilon)$, which has the initial behavior
$D\sim \rev{\varepsilon^{-6/5}}$, but saturates at a constant $\bar{D}$ at
large $\varepsilon$. Likewise, we assume that the surface coverage by
an adsorbed chain behaves like $A_{c}\sim N/g(\varepsilon)$, where
$g(\varepsilon)\sim \rev{\varepsilon^{2/5}}$ for $\varepsilon\to0$, and
$g(\varepsilon)\to\bar{g}$ at large $\varepsilon$. The relation
$\sigma q=\sigma^{\star}$ is taken to be still valid in the saturation
regime, i.e., the thickness of the adsorbed sublayer is constant
throughout the partially adsorbed regime (ii). Repeating the analysis
of Section \ref{sec:scaling_regimes}, we obtain similar results,
except that the crossover points from (i) to (ii) and from (ii) to
(iii) are now given by 
\begin{equation}
(\sigma N/g(\varepsilon))^{*}\sim1\quad\mbox{and}\quad(\sigma D(\varepsilon))_{0}\sim1.
\end{equation}

Regarding the comparison of the theoretical predictions with the SCF
calculations, we find that most of our previous conclusions still
hold: At fixed sufficiently large $\varepsilon$ and $\sigma$, the
height of the brush as a function of $N$ still exhibits the crossover
from constant to linear behavior 
\begin{equation}
H\sim\left\{ \begin{array}{ll}
D\sim \rev{\varepsilon^{-6/5}} & :\;N<N^{\star}
  \sim\rev{\varepsilon^{2/5}}/\sigma\\
N[\sigma-\sigma^{\star}]^{1/3} & :\;N>N^{\star}
\end{array}\right.,
\end{equation}
with a transition point $(\sigma N)^{*}\sim g(\varepsilon)$ that
is independent of $\sigma$. The regime of partial adsorption is even
wider than before: In fact, when increasing $\varepsilon$ for large
$N$, the transition (ii)-(iii) can no longer take place if $\bar{g}<\sigma N$.
The rescaled fraction of adsorbed chains, $q\sigma N\sim g(\varepsilon)$
is still predicted to collapse for different $\sigma$ and $N$. Also,
the density of adsorbed contacts $N\sigma f\frac{1}{D}\sim g(\varepsilon)/D(\varepsilon)$
and the fraction of contacts in the adsorbed phase $\thetaads1/D(\varepsilon)$
are still independent of $N$ and $\sigma$. Both curves are now predicted
to rise and the saturate at $\bar{g}/\bar{D}$ and $1/\bar{D}$, respectively,
in agreement with the SCF data. In particular, the data for $\thetaads$
in Fig.\ \ref{fig:theta in ads phase} suggest that $D(\varepsilon)$
saturates quickly already at $\varepsilon\sim3$, in agreement with
Fig.\ \ref{fig:density profiles}.

\section{Phase transitions of single ''probe'' chains in the brush}

It is commonly understood that polymer adsorption involves a phase
transition. An isolated chain on a planar substrate undergoes a continuous
phase transition (which is smoothed out by finite size effects) and
one can identify the critical \rev{(or, more precisely,
multicritical)} adsorption point in the $N\rightarrow\infty$ limit
\cite{Klushin:2011, Eisenriegler:1982}
A minority adsorption-active chain inserted in a neutral brush
undergoes a much sharper first-order-like transition where the
transition point is affected by the brush density and the relative
lengths of the minority and majority chains
\citep{Skvortsov:1999,Klushin:2014,Qi:2015}.  In the case of the
monodisperse brush where all the chains are adsorption-active we
encounter a strange and counter-intuitive picture. On the one hand, we
have identified microphase separated states which suggests phase
coexistence and some underlying first-order transition. On the other
hand, we see that all the characteristics of the brush as a whole as
well as of individual chains change quite smoothly with increasing
adsorption strength, $\varepsilon$. Phase coexistence is normally
expected to be confined to a line in the pressure-temperature plane.
In our situation, the brush is not loaded, which corresponds to a
fixed zero osmotic pressure. The temperature is associated with the
adsorption parameter, but contrary to naive expectations of a
transition point, we observe coexistence in a broad range of values of
$\varepsilon$. In order to analyze this situation in more detail we
show in Figure \ref{fig:density profiles}b how the total
self-consistent field (including attraction to the surface) changes
with the increase in the adsorption parameter. The field profile
includes the attractive well of the width of one layer and a broad
weakly repulsive barrier. The depth of the attractive well changes
very little and remains in the range of 0.2-0.4 $k_{B}T$ when
$\varepsilon$ increases from 0.4 to 10 $k_{B}T$, see the inset in Fig
\ref{fig:density profiles}b. Simultaneously, the repulsive barrier is
also slightly adjusted. This delicate adjustment allows to maintain
the broad bi-modal distribution of the free end reflecting a
conformation that fluctuates between two phases, as demonstrated
earlier in Figure \ref{fig:end distributions}.

\begin{figure}
\begin{centering}
\includegraphics[width=7cm]{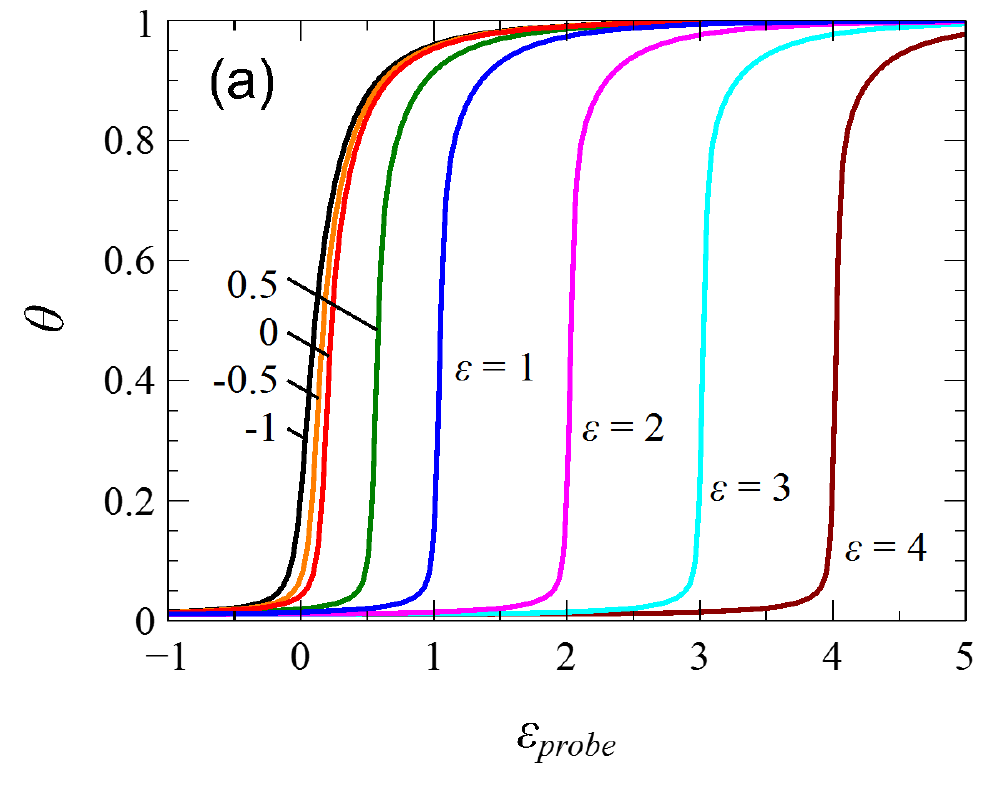} \includegraphics[width=7cm]{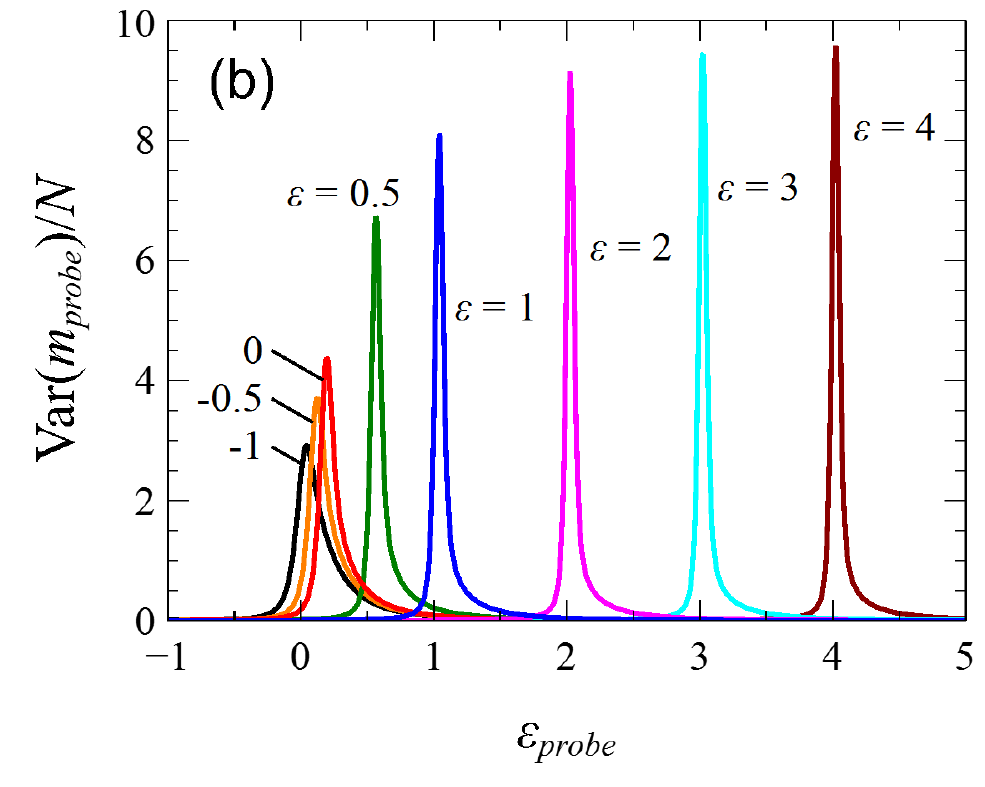}
\par\end{centering}
\caption{\label{fig:theta probe}Average fraction of adsorbed units in the
probe chain (a) and its variance (b) as function of the probe chain
interaction energy $\varepsilon_{probe}$ at various brush chains
interaction energy $\varepsilon,$ as indicated. The
probe chain is inserted in the brush with $N=100$ and $\sigma=0.04$.
The probe chain length is equal to the brush chain length.}
\end{figure}

To gain more insight into this intriguing system we consider
a virtual 
probe chain that differs from all the other brush chains only by its
affinity to the substrate which is taken as a new independent parameter,
$\varepsilon_{probe}$ . We fix all the brush parameters such as $N,\sigma,\varepsilon$,
and study the properties of the probe chain exposed to the fixed brush
potential, as a function of its own adsorption parameter $\varepsilon_{probe}$
. The average fraction of the adsorbed monomers in the probe chain,
$\theta(\varepsilon_{probe})$ is shown in Figure \ref{fig:theta probe}a.

\begin{figure}
\begin{centering}
\includegraphics[width=7cm]{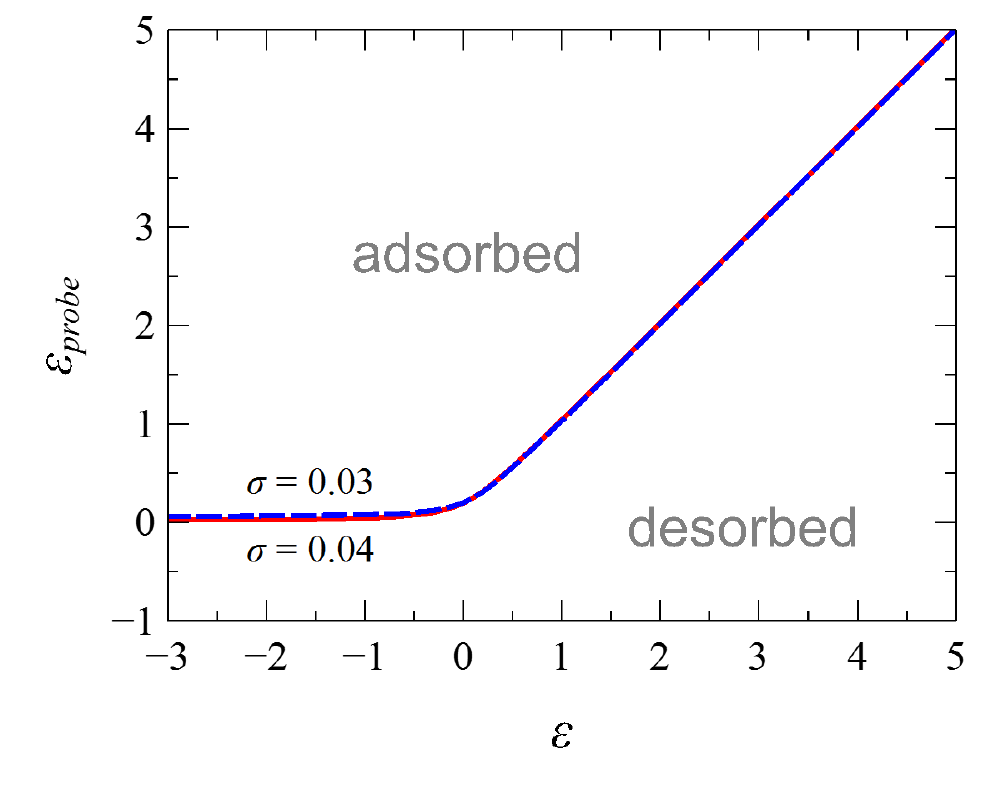}
\par\end{centering}
\caption{\label{fig:Phase-diagram}Phase diagram of a probe chain with its
own adsorption parameter $\varepsilon_{probe}$ in a monodisperse
brush with adsorption parameter $\varepsilon$ . The probe chain is
otherwise the same as the other brush chains.}
\end{figure}

The average fraction of the adsorbed monomers in the probe chain,
$\theta(\varepsilon_{probe})$ is shown in Figure \ref{fig:theta
probe}a.  It is clear that for strong enough attraction of the brush
chains, $\varepsilon\geq0.5$, the number of contacts of the probe
chain sharply increases when its affinity parameter matches that of
the brush itself, $\varepsilon_{probe}\simeq\varepsilon$. The
sharpness of the transition as quantified by the peak value of the
reduced mean-square fluctuations, $Var(m)/N=\left(\langle
m^{2}\rangle-\langle
m\rangle^{2}\right)/N=d\theta/d\varepsilon_{probe},$ see Figure
\ref{fig:theta probe}b. One can see that the sharpness generally
increases with the brush parameter $\varepsilon$ but saturates at
larger $\varepsilon$. The same sharp transition can be seen using the
average height of the free end as an indicator.  We summarize the
properties of the probe chain by presenting the phase diagram in the
$\left(\varepsilon_{probe},\varepsilon\right)$ plane, see Figure
\ref{fig:Phase-diagram}. 

\begin{figure}
\begin{centering}
\includegraphics[width=7cm]{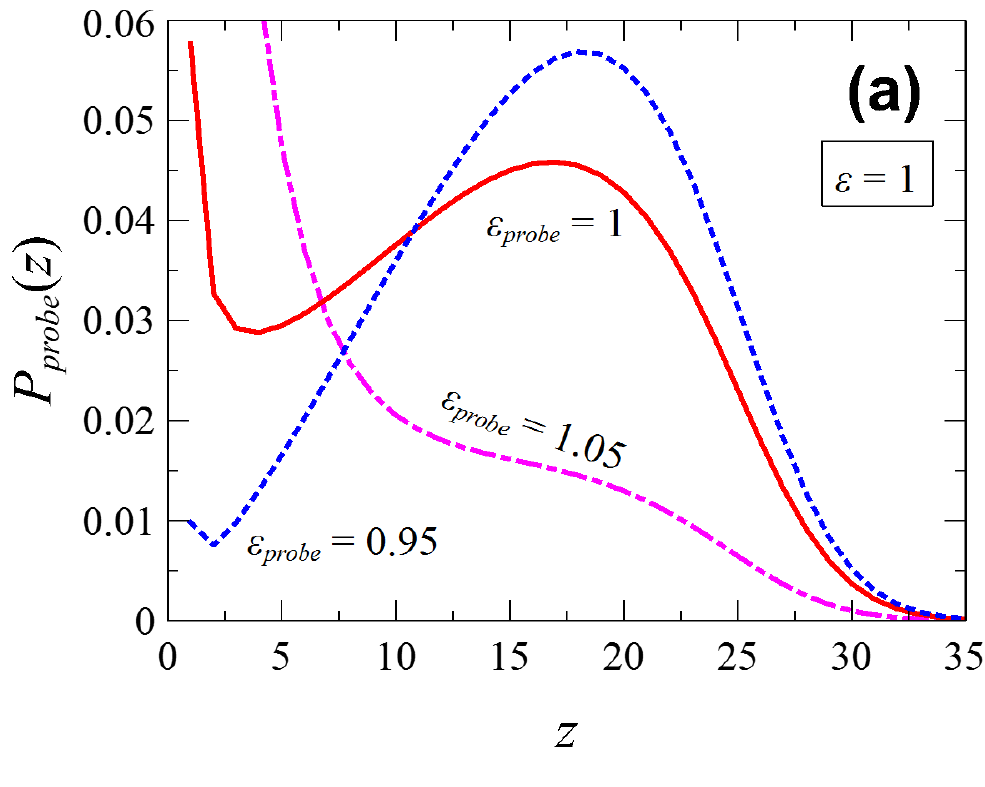} \includegraphics[width=7cm]{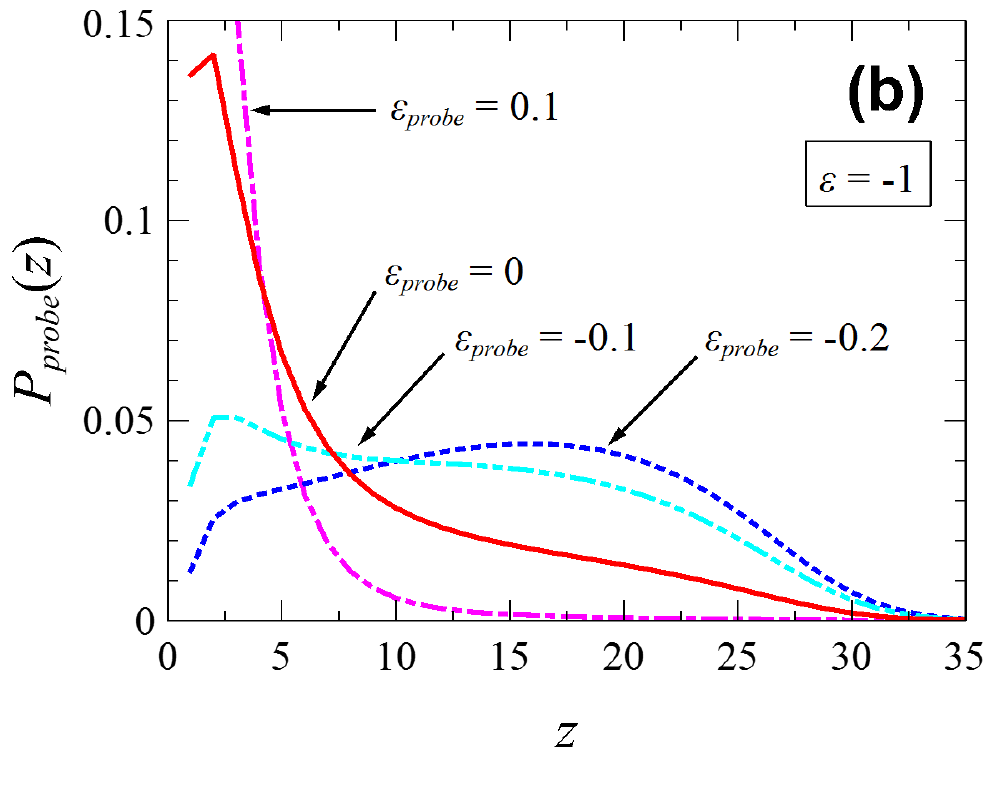}
\par\end{centering}
\caption{\label{fig:probe chain distributions}Distributions $P_{e}(z)$ for
the free ends of the probe chain in a monodisperse brush with $N=100$
and $\sigma=0.04$ at various the probe chain interaction energy $\varepsilon_{probe}$
, as indicated, and brush polymer adsorption energy $\varepsilon=1$
(a) and $-1$ (b). }
\end{figure}

To identify the nature of the underlying transitions we present a
more detailed study for the case of a relatively strongly attractive
substrate, $\varepsilon=1$, and for the case when the brush
chains experience some extra repulsion, $\varepsilon=-1$. We specifically
pay attention to two criteria identifying first-order transitions
in finite systems: 1) bimodal distributions in the vicinity of the
transition point, and 2) the variance of extensive parameters (such
as the number of monomers in contact with the substrate) growing $\varpropto N^{2}$
as consistent with the system fluctuating between two distinct phases.
The changes in the shape of the distribution of the end monomer of
the probe chain upon crossing the line separating the probe chain
phases are displayed in Figure \ref{fig:probe chain distributions}.
It is clear that bi-modality emerges at the transition in
the case of the attractive brush (a), but is absent in the case of
extra repulsion (b). We conclude that part of the phase diagram (at
large positive $\varepsilon$) can be interpreted as a line
of I-order transitions, while the other part (at large negative $\varepsilon$)
resembles more a line of II-order continuous transitions.

\begin{figure}
\begin{centering}
\includegraphics[width=7cm]{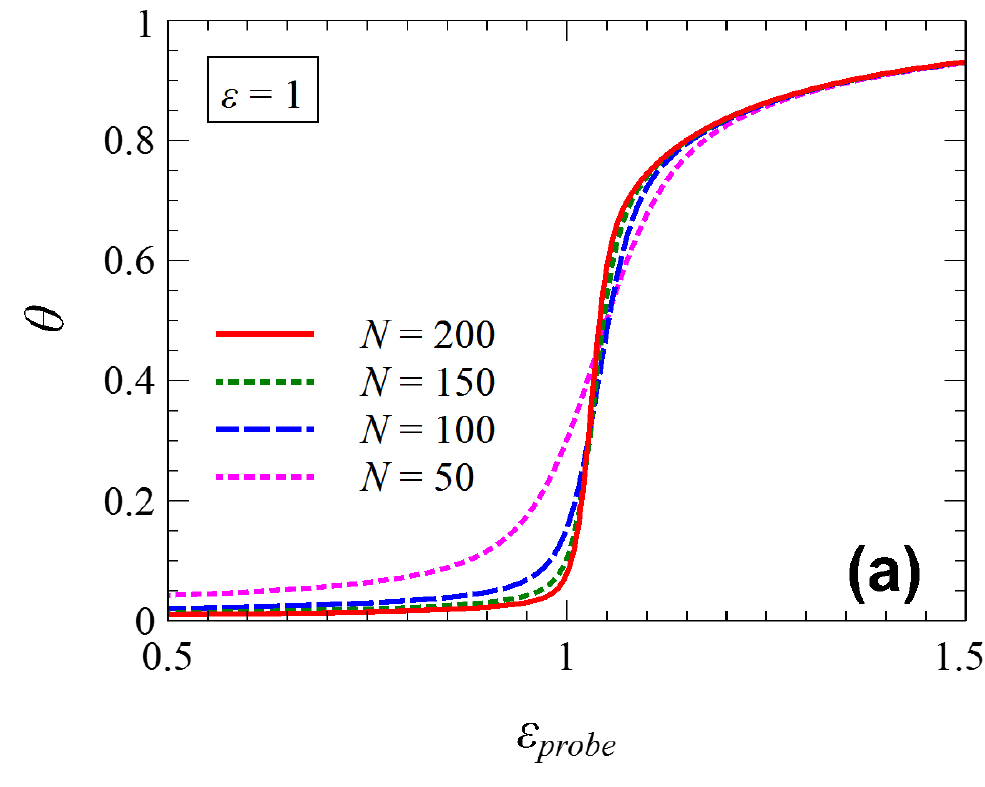} \includegraphics[width=7cm]{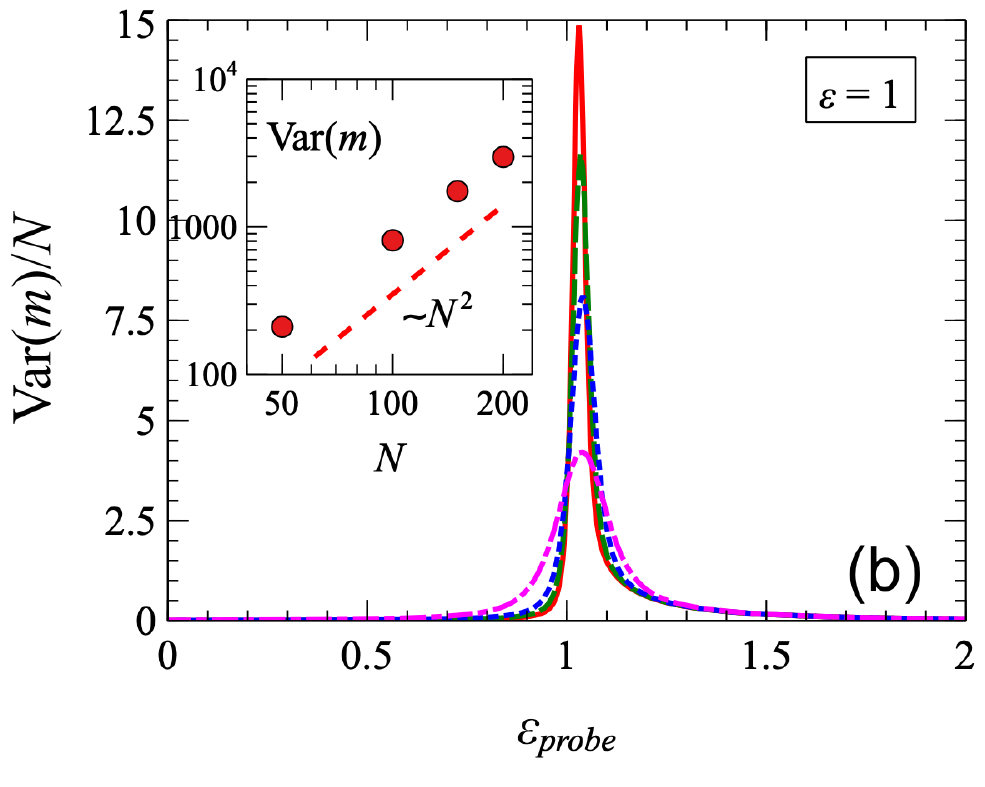}
\par\end{centering}
\caption{\label{fig:finite size I}Fraction of adsorbed units in the probe
chain (a) and its variance (b) as function of the probe chain interaction
energy $\varepsilon_{probe}$ at fixed brush chains interaction energy
$\varepsilon=1$. }
\end{figure}

\begin{figure}
\begin{centering}
\includegraphics[width=7cm]{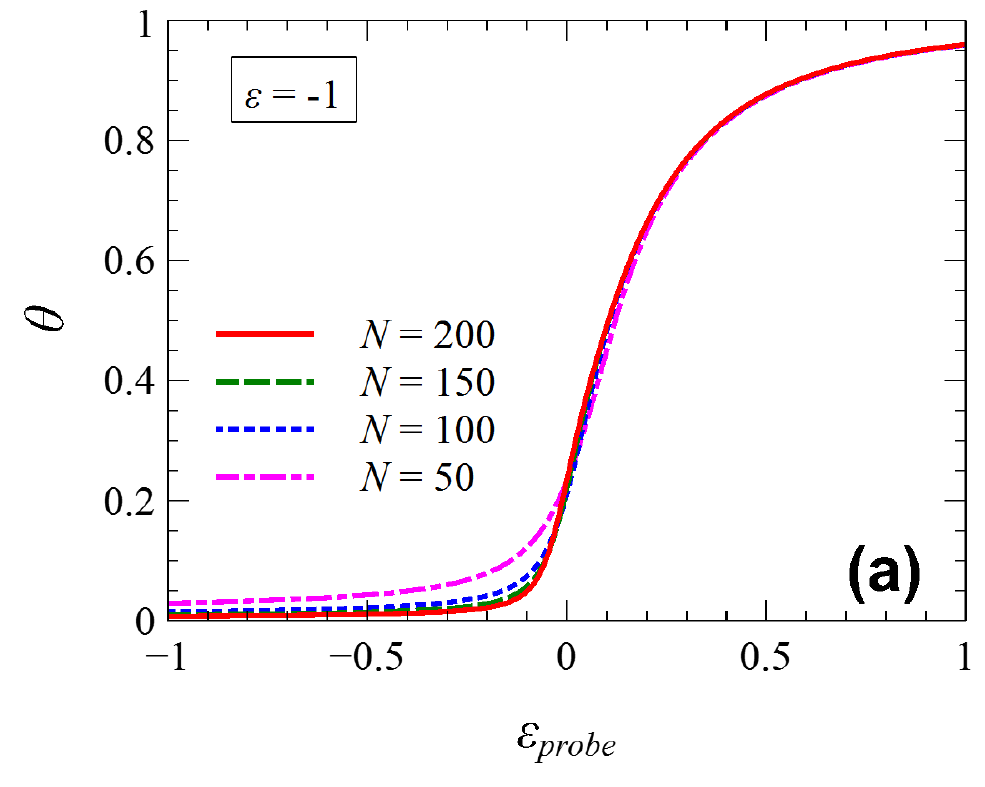} \includegraphics[width=7cm]{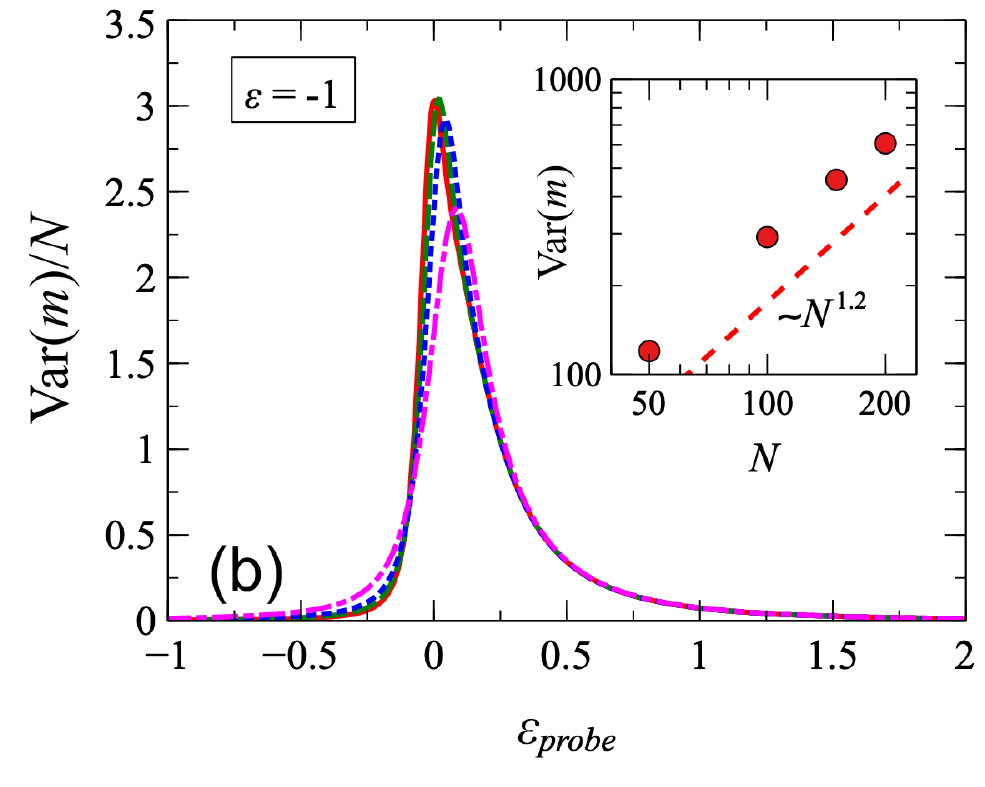}
\par\end{centering}
\caption{\label{fig:finite size II}Fraction of adsorbed units in the probe
chain (a) and its variance (b) as function of the probe chain interaction
energy $\varepsilon_{probe}$ at fixed brush chains interaction energy
$\varepsilon=-1$. }
\end{figure}

To verify this conclusion we study finite chain length effects.
Figure \ref{fig:finite size I}a demonstrates the adsorption curves
of the probe chain in the attractive brush with $\varepsilon=1$ for
several values of the brush and probe chain length $N$ (noting
that both are the same in the present study). 
In order to maintain the same relative magnitude of the adsorption
effect on the brush, we fix the value of the product $N\sigma=4$.
As the system size $N$ increases, the adsorption curves become steeper
although they all intersect at approximately the same point. This
behavior is also typical for finite-size effects in I-order transitions
\citep{Klushin:2011}. The fluctuations near the transition point
become more prominent with increasing $N$, see Figure \ref{fig:finite size I}b,
while the position of the peak is unaffected by the system size. The
inset demonstrates that the peak values scale as $Var(m)_{peak}\propto N^{2}$
confirming the I-order type transition. On the other hand, if the
probe chain undergoes the adsorption transition in a brush with extra
repulsion, $\varepsilon=-1$, the finite chain length effects
look different. The probe chain adsorption curves do not cross
for different values of $N$, see Figure \ref{fig:finite size II}a.
As for the fluctuations in the number of contacts, the position of
the peak shifts to lower values of $\varepsilon_{probe}$ with the
increase in $N$, and the maximum variance scales as $Var(m)_{peak}\propto N^{x}$
with $x\approx1.2$, see Figure \ref{fig:finite size II}b

We conclude that the phase diagram contains a line of I order transitions
$\varepsilon_{probe}=\varepsilon$ for large positive $\varepsilon$.
This line eventually degenerates into a line of II-order transitions
at negative (and, possibly, very small positive) values of $\varepsilon$
where the transition point for the probe chain become almost independent
of brush adsorption parameter. From the general mean-field picture
of phase transitions one would expect a tricritical point where the
lines of the I and II-order transitions are joined together \citep{Chaikin:1995,Klushin:2011}.
The exact localization of this point and a detailed study of
the tricritical adsorption is outside the scope of the present paper.

Going back to a homogeneous \rev{adsorption-active} brush
with a single adsorption parameter $\varepsilon$ we \rev{conclude
that the brush maintains} the density profile exactly corresponding
to the line of I-order transitions, and with increasing $\varepsilon$
we just move up along that line, never leaving it. This peculiar picture
seems to be an inevitable consequence of three conditions: 1) the
total number of monomers exceeds the maximum adsorption capacity of
the substrate (``hairy'' saturation regime) which means that two
states must exist; 2) An adsorption scenario whereby
a certain fraction of chains is adsorbed thus ensuring phase coexistence
at the level of brush chains rather than at the level of monomers;
and 3) brush monodispersity which means that all the chains are identical
and therefore each chain has to fluctuate strongly between the two
phases. 

Yet another puzzle appears when we look at the fluctuations
of the number of adsorbed monomers, $n$, per unit area in a homogeneous
monodisperse brush. The average number of contacts coincides with
the monomer density $\varphi_{1}$. According to standard thermodynamic
formulas, $\varphi_{1}=\left\langle n\right\rangle =-\frac{\partial F}{\partial\varepsilon}|_{N,\sigma}$
where $F$ is the free energy of the brush per unit area. Mean-square
fluctuations in the number of contacts are given by the second derivative,
$Var\left(n\right)=-\frac{\partial^{2}F}{\partial\varepsilon^{2}}|_{N,\sigma}=\frac{\partial\varphi_{1}}{\partial\varepsilon}|_{N,\sigma}$
which is the slope of the adsorption curve for a homogeneous brush.
It is clear from Figure \ref{fig:first layer density} that the slope
is essentially independent of the chain length and does not carry
any indication of the large chain fluctuations. We come to a conclusion
that although a monodisperse brush is composed of strongly
fluctuating chains, these fluctuations must be correlated in such
a way that the brush as a whole represents a regular thermodynamic
system with a perfectly normal fluctuation behavior. Within the SCF
framework it is impossible to verify this picture by following up
the correlated changes in the conformations of the neighboring chains,
so a complete resolution of this fluctuation paradox would requite
a MC of MD simulation.

\section{Discussion and Summary}

We have demonstrated that a monodisperse brush with a strong enough
monomer attraction to the substrate forms a micro-phase separated
system with a fraction of chains being in close contact to the surface
while the rest of the chains form a residual brush with a reduced
effective grafting density. As the adsorption energy $\varepsilon$ is
increased, the fraction of chains in the adsorbed phase initially
increases but eventually the system saturates at large values of
$\varepsilon\gtrsim3$.  Phase coexistence is retained in a very broad
range of $\varepsilon$ and it is impossible to identify a transition
point. Indirectly, the adsorption parameter controls the grafting
density of the residual brush. Depending on the value of the product
$\sigma N$, in the saturation limit the residual brush may disappear
completely or be reduced to isolated mushroom-like tails. We have
identified and studied the regime, $\sigma N>2.5$ in which the
residual brush is reasonably well defined for any value of the
adsorption energy $\varepsilon$.

In some respect, our observations described above are reminiscent
of partial wetting \cite{deGennes:1985} as can be observed, e.g.,
in a gas in contact with an attractive substrate if one approaches
the gas/liquid transition: A thin liquid layer forms on the substrate
for a wide range of adsorption strengths, similar to the adsorption
layer in our adsorption-active brushes. In other respect, however,
the situation here is very different from regular wetting: The attractive
substrate not only influences the thickness of the wetting layer,
but also the properties of the ''coexisting'' phase, e.g., the effective
grafting density of the outer brush. The connectivity of the brush
chains transmits a strong coupling between the adsorbed and desorbed
layer. It is reflected in strong fluctuations of individual chains
between ''coexisting'' adsorbed desorbed states, which are not sharply
defined as the free energy barrier separating them is always small,
less than 1$kT$ even when the adsorption energy is as large as $\varepsilon=10$.
This means that kinetic trapping is most likely absent and the exchange
of monomers between phases must be characterized by relatively fast
dynamics. Overall, certain features of the behavior of the adsorption-active
brush are rather difficult to fit in the conventional framework of
the phase transition theory. We have attempted to clarify the situation
by introducing the notion of a probe chain and by constructing its
phase diagram. This construction answers some questions but raises
several others.

\rev{A coexistence between stretched and collapsed chains or chain
parts has been observed before in polyelectrolyte systems
~\cite{Wang:2006}. In the case of polyelectrolyte brushes grafted on
oppositely charged substrates the situation is very similar to the one
considered here. The the charged substrate provides the attractive
part of the potential, which is neutralized by the combination of
electrostatic screening and steric repulsion due to the dense proximal
layer, while the rest of the chains form a residual brush governed by
repulsive mean force~\cite{Merlitz:2015}
Another possibility involves an electrically neutral substrate covered
by a polyamphiphylic brush with short chains of one charge and much
longer chains of the opposing charge~\cite{Shusharina:2001}
Here the short blocks create an effective attractive potential localized
near the substrate which again competes with a longer-range repulsion.
In both cases, a bimodal distribution of global chain characteristics
was observed indicating phase coexistence along the second scenario.
A more intricate scenario suggesting lateral segregation into collapsed
and stretched microphases was observed for weak polyelectrolyte brushes
in a poor solvent~\cite{Tagliazucchi:2010}
}

In the present study, we have discussed monodisperse polymer
brushes only. Polydispersity effects will be investigated in future
work. Based on previous results on polydispersity effects in responsive
brushes \cite{Qi:2018}, we expect that the main features of the adsorption-active
brush reported above will persist, although the fluctation characteristics
of single chains will likely change \cite{Qi:2016}. 
Although our discussion was confined to the case of a brush formed
by flexible uncharged chains grafted onto a planar substrate and immersed
in a good solvent, we believe that the qualitative results concerning
the microphase separation and the indirect control of the residual
brush by the adsorption parameter are applicable to a much broader
class of situations. 

The data that support the findings of this study are available from
the authors upon reasonable request.

\section{Dedication}
This paper is dedicated to Tatiana Birshtein who made an outstanding
contribution to the contemporary statistical physics of macromolecules
and to a great extent promoted the progress in theoretical and experimental
polymer science worldwide.

\begin{acknowledgments}
Financial supported by the Russian Foundation for Basic Research through
the grant no. 20-53-12020 NNIO\_a and by the German Science Foundation
through the grant Schm 985/23 is gratefully acknowledged. 
\end{acknowledgments}

\appendix

\section{The Scheutjens-Fleer self-consistent field method}

In the case of a planar brush, we use a one-gradient version of the
SF-SCF method, in which the lattice sites are organized in a planar
layers, each layer is referred to with a coordinate $z$ normal to
the grafting plane. \rev{The model is limited to laterally
homogeneous systems. In framework of the SF-SCF method, various interactions
between the particles in the system are replaced by the mean effective
interactions, or the potential $u(z)$. Hence, SCF does not exclude
overlapped (not self-avoiding) conformations. In principle, this may
result in an underestimation of the entropic cost of collapsing the
chains to the attractive surface. However, a detailed study of adsorbed
polymer layers has demonstrated an excellent match of the SCF and
simulation results~\cite{Fleer:1998}.}

If the potential $u(z)$ is specified, then the statistical weight
of a monomer unit is $G(z)=\exp[-u(z)]$. Using the monomer unit's
weight, we calculate two statistical weights: the statistical weight
$G_{t}(z,s)$ of a chain with $s$ monomer units tethered at one end
at the surface and having its other end in the layer $z$ (the subscript
``\textit{t}'' means ``tethered'') and the the statistical weight
$G_{f}(z,s)$ of a chain with $s$ monomer units having one end pinned
in the layer $z$ and the other end free (the subscript ``\textit{f}''
means ``free''). They satisfy the recurrence relations

\begin{equation}
G_{t,\,f}(z,s+1)=G(z)[\lambda G_{t,\,f}(z-1,s)+(1-2\lambda)G_{t,\,f}(z,s)+\lambda G_{t,\,f}(z+1,s)].\label{eq:Gtf}
\end{equation}
where $\lambda$ is the probability that a random walk step connects
neighboring layers. On the simple cubic lattice, one has $\lambda=1/6$.
The initial condition for $G_{t}(z,s)$ and $G_{f}(z,s)$ are different

\begin{equation}
G_{t}(z,1)=\begin{cases}
G(z), & z=1\\
0, & z\neq1
\end{cases}
\end{equation}
and

\begin{equation}
G_{f}(z,N)=\begin{cases}
G(z), & z\geq1\\
0, & z<1
\end{cases}
\end{equation}
The recurrence relation (\ref{eq:Gtf}) should be modified in the
first layer adjacent to the grafting surface: for $z<1$, $G_{t,\,f}(z,s)=0$,
hence

\begin{equation}
G_{t,\,f}(1,s+1)=G(z)[(1-2\lambda)G_{t,\,f}(1,s)+\lambda G_{t,\,f}(2,s)].\label{eq:Gtf_BC}
\end{equation}
This plays the role of the boundary condition at $z=1$. It is also
obvious that for $z>s$, $G_{t}(z>s,s)=0$.

By using the set of $G_{t}(z,s)$ and $G_{f}(z,s)$ for $1\leq s\leq N$
one can calculate the polymer volume density profile via the composition
law \citep{Fleer:1993}:

\begin{equation}
\varphi(z)=\frac{\sigma}{q}\cdot\frac{\sum_{s=1}^{N}G_{t}(z,s)G_{f}(z,N-s+1)}{G(z)}.\label{eq:phi}
\end{equation}

The factor $G(z)$ in the denominator of the rhs of (\ref{eq:phi})
is used to avoid double counting of the monomer unit in the layer
$z$. The normalization constant is obtained from the condition $\sum_{z}\varphi(z)=N\sigma$
and includes the partition function of a tethered chain $q=\sum_{z}G_{t}(z,N)=G_{f}(1,N)$.

At each lattice layer, the incompressibility condition is obeyed:
\begin{equation}
\varphi(z)+\varphi_{s}(z)=1.\label{eq:incompressibility}
\end{equation}

The potential $u(z)$ acting on the monomer units

\begin{equation}
u(z)=-\log[1-\varphi(z)]-\varepsilon\delta_{z,1}\label{eq:u}
\end{equation}
where the second term is the additional surface attraction energy
in the first lattice layer ($z=1$). $\varepsilon$ is the measure
of the polymer-surface attraction, $\delta_{i,j}$ is the Kronecker
delta.

The system of equations Eqs. (\ref{eq:Gtf}), (\ref{eq:Gtf_BC}),
(\ref{eq:phi}), and (\ref{eq:u}) is solved self-consistently taking
into account the incompressibility condition (\ref{eq:incompressibility}).
That is, with an initial guess $u(z)$ one calculates the set of propagators
$G_{t}(z,s)$ and $G_{f}(z,s)$ {[}Eqs. (\ref{eq:Gtf}) and (\ref{eq:Gtf_BC}){]},
the density profile $\text{\ensuremath{\varphi(z)}}$ {[}Eq. (\ref{eq:u}){]}
and the new field $u(z)$ {[}Eq. (\ref{eq:u}){]}. The procedure is
then iteratively repeated until it converges to a fixed point, or
the self-consistent solution. Once the solution, i.e the self-consistent
potential $u(z)$ is found, this gives access to the density profile,
end segment distribution, and can also be used to study the behavior
of a probe minority chain inserted into the brush.

\bibliographystyle{unsrt}
\bibliography{refs}

\begin{thebibliography}{10}

\bibitem{Currie:2003}
E.~P.~K. Currie, W.~Norde, and M.~A. Cohen~Stuart.
\newblock {Tethered polymer chains: surface chemistry and their impact on
  colloidal and surface properties}.
\newblock {\em Adv Colloid Interface Sci}, 100-102:205--265, 2003.

\bibitem{Ayres:2010}
N.~Ayres.
\newblock Polymer brushes: Applications in biomaterials and nanotechnology.
\newblock {\em Polym. Chem.}, 1:769--777, 2010.

\bibitem{Urban:2011}
M.~W. Urban.
\newblock {\em {Handbook of Stimuli-Responsive Materials}}.
\newblock Wiley-VCH Verlag GmbH \& Co. KGaA, Weinheim, Germany, 2011.

\bibitem{Jaquet:2013}
B.~Jaquet, D.~Wei, B.~Reck, F.~Reinhold, X.~Zhang, H.~Wu, and M.~Morbidelli.
\newblock Stabilization of polymer colloid dispersions with ph-sensitive
  poly-acrylic acid brushes.
\newblock {\em Colloid Polym. Sci.}, 291(7):1659--1667, 2013.

\bibitem{Motornov:2003}
M.~Motornov, S.~Minko, K.-J. Eichhorn, M.~Nitschke, F~Simon, and M~Stamm.
\newblock Reversible tuning of wetting behavior of polymer surface with
  responsive polymer brushes.
\newblock {\em Langmuir}, 19(19):8077--8085, 2003.

\bibitem{Cohen-Stuart:2010}
M.~A. Cohen~Stuart, W.~T.~S. Huck, J.~Genzer, M.~M\"{u}ller, C.~Ober, M.~Stamm,
  G.~B. Sukhorukov, I.~Szleifer, V.~V. Tsukruk, M.~Urban, F.~Winnik,
  S.~Zauscher, I.~Luzinov, and S.~Minko.
\newblock Emerging applications of stimuli-responsive polymer materials.
\newblock {\em Nat. Mater.}, 9(2):101--113, 2010.

\bibitem{Qi:2015}
S.~Qi, L.~I. Klushin, A.~M. Skvortsov, A.~A. Polotsky, and F.~Schmid.
\newblock Stimuli-responsive brushes with active minority components: {M}onte
  {C}arlo study and analytical theory.
\newblock {\em Macromolecules}, 48(11):3775--3787, 2015.

\bibitem{Chen:2010}
T.~Chen, R.~Ferris, J.~Zhang, R.~Ducker, and S.~Zauscher.
\newblock Stimulus-responsive polymer brushes on surfaces: Transduction
  mechanisms and applications.
\newblock {\em Prog. Polym. Sci.}, 35(1-2):94 -- 112, 2010.
\newblock Special Issue on Stimuli-Responsive Materials.

\bibitem{Gupta:2008}
S.~Gupta, M.~Agrawal, P.~Uhlmann, F.~Simon, U.~Oertel, and M.~Stamm.
\newblock Gold nanoparticles immobilized on stimuli responsive polymer brushes
  as nanosensors.
\newblock {\em Macromolecules}, 41(21):8152--8158, 2008.

\bibitem{Merlitz:2009}
H.~Merlitz, G.-L. He, J.-U. Sommer, and Ch.-X. Wu.
\newblock {Reversibly Switchable Polymer Brushes with Hydrophobic/Hydrophilic
  Behavior: A Langevin Dynamics Study}.
\newblock {\em Macromolecules}, 42(1):445--451, 2009.

\bibitem{Pasch:2014}
H.~Pasch and B.~Trathnigg.
\newblock Multidimensional {HPLC} of polymers.
\newblock 2013.

\bibitem{Fleer:1993}
G.~J. Fleer, M.~A. Cohen~Stuart, J.~M. H.~M. Scheutjens, T.~Cosgrove, and
  B.~Vincent.
\newblock {\em {Polymers at Interfaces}}.
\newblock Chapman and Hall, London, 1993.

\bibitem{Halperin:1994}
A.~Halperin.
\newblock {On Polymer Brushes and Blobology: An Introduction}.
\newblock In Y.~Rabin and R.~Bruinsma, editors, {\em {Soft order in Physical
  Systems. NATO ASI Series}}, volume 323, pages 33--56. Springer, Berlin
  Heidelberg, 1994.

\bibitem{Rubinstein_book}
M.~Rubinstein and R.~H. Colby.
\newblock {\em Polymer Physics}.
\newblock Oxford University Press, Oxford, 2003.

\bibitem{Milner:1988}
S.~T. Milner, T.~A. Witten, and M.~E. Cates.
\newblock Theory of the grafted polymer brush.
\newblock {\em Macromolecules}, 21(8):2610--2619, 1988.

\bibitem{Zhulina:1989}
Ye.~B. Zhulina, V.A. Pryamitsyn, and O.~V. Borisov.
\newblock Structure and conformational transitions in grafted polymer chain
  layers. a new theory.
\newblock {\em Polymer Science U.S.S.R.}, 31(1):205 -- 216, 1989.

\bibitem{deGennes:1976}
P.G. De~Gennes.
\newblock Scaling theory of polymer adsorption.
\newblock {\em J. Phys. France}, 37(12):1445--1452, 1976.

\bibitem{deGennes:1981}
P.~G. De~Gennes.
\newblock Polymer solutions near an interface. adsorption and depletion layers.
\newblock {\em Macromolecules}, 14(6):1637--1644, 1981.

\bibitem{Bouchaud:1987}
E.~Bouchaud and M.~Daoud.
\newblock Polymer adsorption: concentration effects.
\newblock {\em J. Phys. France}, 48(11):1991--2000, 1987.

\bibitem{Descas:2006}
R.~Descas, J.-U. Sommer, and A.~Blumen.
\newblock Concentration and saturation effects of tethered polymer chains on
  adsorbing surfaces.
\newblock {\em The Journal of Chemical Physics}, 125(21):214702, 2006.

\bibitem{Descas:2004}
R.~Descas, J.-U. Sommer, and A.~Blumen.
\newblock Static and dynamic properties of tethered chains at adsorbing
  surfaces: A {M}onte {C}arlo study.
\newblock {\em The Journal of Chemical Physics}, 120(18):8831--8840, 2004.

\bibitem{deGennes:1983b}
P.~G. de~Gennes and P.~Pincus.
\newblock {Scaling theory of polymer adsorption: Proximal exponent}.
\newblock {\em J. Physique - Lettres}, 44:L--241--L--246, 1983.

\bibitem{Grassberger:2005}
P.~Grassberger.
\newblock Simulations of grafted polymers in a good solvent.
\newblock {\em J. Phys. A: Math. Gen.}, 38:323, 2005.

\bibitem{Klushin:2013}
L.~I. Klushin, A.~A. Polotsky, H.-P. Hsu, D.~A. Markelov, K.~Binder, and A.~M.
  Skvortsov.
\newblock Adsorption of a single polymer chain on a surface: Effects of the
  potential range.
\newblock {\em Phys. Rev. E}, 87:022604, Feb 2013.

\bibitem{Zhang:2018}
S.~Zhang, S.~Qi, L.~I. Klushin, A.~M. Skvortsov, D.~Yan, and F.~Schmid.
\newblock Phase transitions in single macromolecules: Loop-stretch transition
  versus loop adsorption transition in end-grafted polymer chains.
\newblock {\em J. Chem. Phys.}, 148:044903, 2018.

\bibitem{Klushin:2011}
L.~I. Klushin and A.~M. Skvortsov.
\newblock Unconventional phase transitions in a constrained single polymer
  chain.
\newblock {\em J. Phys. A: Math. Theor.}, 44(47):473001, 2011.

\bibitem{Eisenriegler:1982}
E.~Eisenriegler, K.~Kremer, and K.~Binder.
\newblock Adsorption of polymer chains at surfaces: Scaling and monte carlo
  analyses.
\newblock {\em The Journal of Chemical Physics}, 77(12):6296--6320, 1982.

\bibitem{Skvortsov:1999}
A.~M. Skvortsov, A.~A. Gorbunov, F.~A.~M.. Leermakers, and G.~J. Fleer.
\newblock Long minority chains in a polymer brush: A first-order adsorption
  transition.
\newblock {\em Macromolecules}, 32(6):2004--2015, 1999.

\bibitem{Klushin:2014}
L.~I. Klushin, A.~M. Skvortsov, A.~A. Polotsky, S.~Qi, and F.~Schmid.
\newblock Sharp and fast: Sensors and switches based on polymer brushes with
  adsorption-active minority chains.
\newblock {\em Phys. Rev. Lett.}, 113:068303, Aug 2014.

\bibitem{Chaikin:1995}
P.~M. Chaikin and T.~C. Lubensky.
\newblock {\em Principles of Condensed Matter Physics}.
\newblock Cambridge University Press, 1995.

\bibitem{deGennes:1985}
P.~G. de~Gennes.
\newblock Wetting: Statics and dynamics.
\newblock {\em Rev. Mod. Physics}, 57:827--863, 1985.

\bibitem{Wang:2006}
Q.~Wang.
\newblock {Modelling Layer-by-Layer Assembly of Flexible Polyelectrolytes}.
\newblock {\em J. Phys. Chem. B}, 110:5825--5828, 2006.

\bibitem{Merlitz:2015}
H.~Merlitz, C.~Li, C.~Wu, and J.-U. Sommer.
\newblock Polyelectrolyte brushes in external fields: molecular dynamics
  simulations and mean-field theory.
\newblock {\em Soft Matter}, 11:5688--5696, 2015.

\bibitem{Shusharina:2001}
N.~P. Shusharina and P.~Linse.
\newblock Oppositely charged polyelectrolytes grafted onto planar surface:
  Mean-field lattice theory.
\newblock {\em Eur. Phys. J. E}, 6(2):147--155, 2001.

\bibitem{Tagliazucchi:2010}
M.~Tagliazucchi, M.~Olvera de~la Cruz, and I.~Szleifer.
\newblock Self-organization of grafted polyelectrolyte layers via the coupling
  of chemical equilibrium and physical interactions.
\newblock {\em PNAS, Proc. Natl. Acad. Sci. USA}, 107(12):5300--5305, 2010.

\bibitem{Qi:2018}
S.~Qi, L.~I. Klushin, A.~M. Skvortsov, M.~Liu, J.~Zhou, and F.~Schmid.
\newblock Tuning transition properties of stimuli-responsive brushes by
  polydispersity.
\newblock {\em Adv. Functional Materials}, 28:1800745, 2018.

\bibitem{Qi:2016}
S.~Qi, L.~I. Klushin, A.~M. Skvortsov, and F.~Schmid.
\newblock Polydisperse polymer brush: Internal structure, critical behavior,
  and interaction with flow.
\newblock {\em Macromolecules}, 49:9665--9683, 2016.

\bibitem{Fleer:1998}
G.~J. Fleer and F.~A.~M. Leermakers.
\newblock Statistical thermodynamics of polymer layers.
\newblock {\em Curr. Opin. Coll. Interf. Sci.}, 2(3):308 -- 314, 1997.

\end{thebibliography}

\end{document}